\documentclass{iopart}
\pdfoutput=1
\usepackage{euscript,iopams,amssymb,amsfonts,graphicx,bm}
\usepackage{setstack}

  \usepackage{epstopdf}
  \usepackage{graphics}

\usepackage{color}

\eqnobysec

\newcommand{\neb}{\bar{\branch}} 

\newcommand{\targT}{\mathcal{X}} 
\newcommand{\branch}{\mathcal{I}}
\newcommand{\flx}{\mathcal{J}}

\newcommand{\uoo}{\psi}

\renewcommand{\e}{\mathrm{e}}

\newcommand{\tu}{\widetilde{p}}

\renewcommand{\P}{\mathbb{P}}

\newcommand{\E}{{\mathbb E}}

\newcommand{\F}{\mathcal F}

\begin{document}

\title[Drift-diffusion on a Cayley tree]{Drift-diffusion on a Cayley tree with stochastic resetting: the localization-delocalization transition.}

\author{Paul C. Bressloff}
\address{Department of Mathematics, University of Utah 155 South 1400 East, Salt Lake City, UT 84112}

\begin{abstract}
In this paper we develop the theory of drift-diffusion on a semi-infinite Cayley tree with stochastic resetting. In the case of a homogeneous tree with a closed terminal node and no resetting, it is known that the system undergoes a classical localization-delocalization (LD) transition at a critical mean velocity $v_c= -(D/L)\ln (z-1)$ where $D$ is the diffusivity, $L$ is the branch length and $z$ is the coordination number of the tree. If $v <v_c$ then the steady state concentration at the terminal node is non-zero (drift-dominated localized state), whereas it is zero when $v >v_c $ (diffusion-dominated delocalized state). This is equivalent to the transition between recurrent and transient transport on the tree, with the mean first passage time (MFPT) to be absorbed by an open terminal node switching from a finite value to infinity. Here we show how the LD transition provides a basic framework for understanding analogous phase transitions in optimal resetting rates. First, we establish the existence of an optimal resetting rate $r^{**}(z)$ that maximizes the steady-state solution at a closed terminal node. In addition, we show that there is a phase transition at a critical velocity $v_c^{**}(z)$ such that $r^{**}>0$ for $v>v_c^{**}$ and $r^{**}=0$ for $v<v_c^{**}$.  We then identify a critical velocity $v^*(z)$ for a phase transition in a second optimal resetting rate $r^*$ that minimizes the MFPT to be absorbed by an open terminal node. Previous results for the semi-infinite line are recovered on setting $z=2$. The critical velocity of the LD transition provides an upper bound for the other critical velocities such that $v_c^*(z)<v_c^{**}(z)<v_c(z)$ for all finite $z$. Only $v_c(z)$ has a simple universal dependence on the coordination number $z$. We end by considering the combined effects of quenched disorder and stochastic resetting.

\newpage

\end{abstract}

\maketitle

\section{Introduction}

Drift-diffusion processes on trees have a wide range of applications in biological physics, ranging from the active transport of vesicles in the dendritic branches of a neuron \cite{Newby09} to the distribution of oxygen through a branched network of bronchiole tubes during respiration \cite{Chang88,Grebenkov05}. There are also a variety of physical processes for which the
underlying topology is tree--like in nature. Typical examples are diffusion limited aggregation, viscous fingering and invasion
percolation \cite{Havlin,Georges}. Tree-like topologies are also of interest from a mathematical perspective, since they are simpler to analyze compared to a study of the same process defined on a regular lattice. This permits investigations of generic features of interest that can
also, in certain cases, be directly relevant to the regular lattice problem in some appropriate limit. For example, it is well
known that Cayley trees and Bethe lattices provide insights into the behavior of various processes on both
infinite--dimensional lattices and finite--dimensional lattices in the mean field limit \cite{Ziman,Baxter}. 

Recently, we analyzed a first passage time (FPT) problem for drift-diffusion on a homogeneous semi-infinite Cayley tree with an absorbing target at the primary or terminal node of the tree \cite{Bressloff21}. (At the level of a single particle, the drift-diffusion equation is equivalent to a Fokker-Planck (FP) equation.) 
We assumed that the drift velocity $v$, diffusivity $D$, and
branch length $L$ were the same in each branch so that we could exploit the recursive nature of the semi-infinite tree. We calculated the Laplace transform of the flux through the target using an iterative method developed in Ref. \cite{Newby09}, which was used to determine the hitting probability and mean FPT (MFPT). In particular, we showed that for a given coordination number $z$, there is a second-order phase transition from recurrent (drift-dominated) to transient (diffusion-dominated) transport at a critical velocity $v=v_c(z)=-(D/L)\ln (z-1)$. That is, the hitting probability $\pi=1$ and the MFPT $T<\infty$ when $v<v_c$, whereas $\pi <1$ and $T=\infty$ when $v>v_c$. In terms of the P\'{e}clet number Pe $= vL/D$, the phase transition occurs at the critical point Pe $= -\ln(z-1)$
This transition point is identical to a classical localization-delocalization (LD) threshold for the steady-state solution on a tree with a closed rather than an open terminal node \cite{Bressloff96,Bressloff97}. An initial concentration localized at the terminal node tends to diffuse away from the origin, but is counteracted by an inward velocity field on the tree. If, in steady state, the concentration remaining
at the origin has not decayed to zero, the
system is localized, otherwise it is delocalized. Moreover, the LD transition occurs at  $v=v_c $ \cite{Bressloff96,Bressloff97}.

In this paper we further develop the theory of drift-diffusion on a Cayley tree by considering the effects of stochastic resetting \cite{Evans11a,Evans11b}. Prior to absorption at the terminal node, a particle may instantaneously reset to the initial position $x_0$ at a random sequence of times generated by an exponential probability density $\psi(\tau)=r\e^{-r\tau}$, where $r$ is the resetting rate. (One could also incorporate resetting delays -- finite return times and refractory periods - and non-exponential resetting statistics \cite{Mendez19,Evans19a,Mendez19a,Bodrova20,Pal20,Bressloff20A,Evans20}. However, we focus on the simplest case here.) Drift-diffusion with stochastic resetting has recently been studied in the case of the semi-infinite line \cite{Ray19}, which is equivalent to a homogeneous Cayley tree with coordination number $z=2$. In particular, they identified a second-order phase transition involving the optimal resetting rate $r^*$ at which the MFPT is minimized. They showed analytically that there was a sharp transition between a diffusion-dominated regime and a drift-dominated regime at a critical velocity $v_c^*$ such that $r^*>0$ for $v>v_c^*$ and $r^*=0$ for $v<v_c^*$. Since the branch length $L$ plays no role when $z=2$, at least in the homogeneous case, the phase transition can be characterized in terms of another P\'{e}clet number $\mbox{Pe}'=vx_0/D$, where $x_0$ is the distance of the target from the origin. The critical velocity $v_c^*$ occurs when $\mbox{Pe}'=-2$. Comparison with the LD transition for $z=2$, $\mbox{Pe}=0$, it follows that $v_c^* <v_c=0$.\footnote{In Ref. \cite{Ray19} the target is taken to be at $x=L$ and the reset point is at the boundary $x=0$. Therefore, the results of \cite{Ray19} are obtained by reversing the sign of $v$.} The phase transition was further analyzed in \cite{Pal19} as part of a more general study regarding phase transitions in optimal resetting rates.

The main goal of the current paper is to show how the LD transition for drift-diffusion on a Cayley tree provides a basic framework for understanding analogous phase transitions in optimal resetting rates, including the optimal resetting rate $r^*$ considered in \cite{Ray19,Pal19} for $z=2$. We show that $v_c(z)$ is an upper bound for the critical velocities associated with these other phase transitions, and that only $v_c(z)$ has a simple universal dependence on the coordination number $z$. The structure of the paper is as follows. In Sect. 2 we introduce the basic model and construct the general solution in Laplace space. The steady-state solution with stochastic resetting and a closed terminal node is considered in Sect. 3. In particular, we use the results of Sect. 2 to determine how the steady state solution at the closed terminal node depends on the resetting rate $r$ and explore the the emergence of a sharp LD transition in the limit $r\rightarrow 0^+$. We also establish the existence of an optimal resetting rate $r^{**}$ that maximizes the steady-state solution at the terminal node. Moreover, there is a phase transition at a critical velocity $v_c^{**}(z)$ such that $r^{**}>0$ for $v>v_c^{**}$ and $r^{**}=0$ for $v<v_c^{**}$. If $z=2$, the the critical velocity occurs when $\mbox{Pe}'=-1$. In Sect. 4 we develop the corresponding analysis of the hitting probability and MFPT at an open terminal node, recovering the results of \cite{Ray19,Pal19} when $z=2$. Although we focus on the case of a homogeneous tree in this paper, possible extensions to quenched disorder are discussed in Sect. 5.

\setcounter{equation}{0}
\section{Drift-diffusion equation on a Cayley tree}  

\begin{figure}[b!]
  \raggedleft
  \includegraphics[width=13cm]{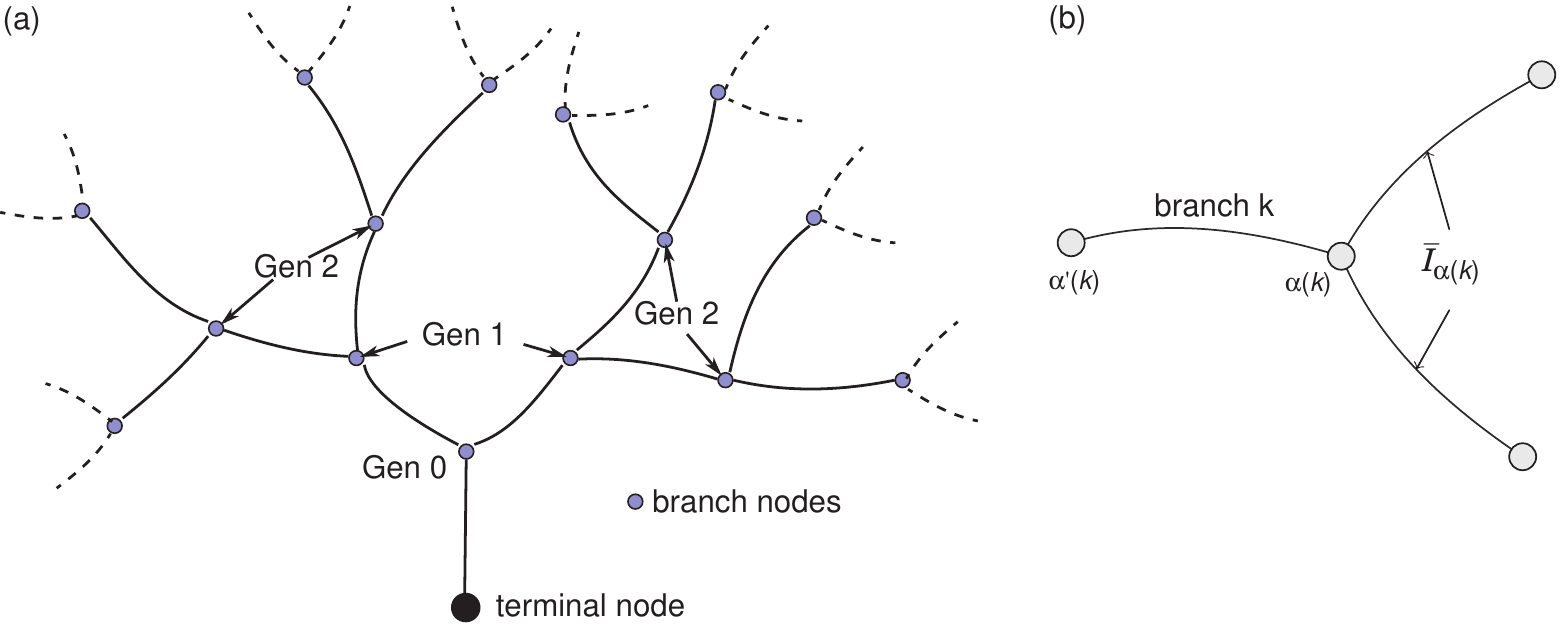}
  \caption{(a) Semi-infinite Cayley tree with coordination number $z=3$ and either an absorbing or reflecting boundary at the primary node. (b) A branch node
    $\alpha(k)$ is shown in relation to the neighboring branch
    node  $\alpha'(k)$ closest to the primary or terminal node.  The
    branch segments extending out from $\alpha(k)$ in the positive
    direction together comprise the set $\neb_{\alpha(k)}$.  }
  \label{fig:notation}
\end{figure}

Following \cite{Bressloff21}, we begin by considering an drift diffusion equation on a homogeneous, semi-infinite Cayley tree $\Gamma$ with coordination number $z$ and either an absorbing (open) or reflecting (closed) boundary at the terminal (primary) node of the tree. The example of $z=3$ is shown in Fig. \ref{fig:notation}(a). In order to write down the drift-diffusion equation on each branch, it is useful to introduce a convenient labeling scheme for the nodes and branches. 
Let $\alpha_{0}$ denote the first branch node opposite the terminal node. For every other branching node $\alpha \in
\Gamma$ there exists a unique direct path from $\alpha_0$ to
$\alpha$ (one that does not traverse any line segment more than
once). We can label each node $\alpha \neq \alpha_0$ uniquely by the
index $k$ of the final segment of the direct path from $\alpha_0$ to
$\alpha$ so that the branch node corresponding to a given 
segment label $k$ can be written $\alpha(k)$. We denote the other node of
segment $k$ by $\alpha'(k)$. Taking the primary branch to be $k=0$, it follows that $\alpha(0)=\alpha_0$ and $\alpha'(0)$ is the terminal node. We also
introduce a direction on each segment of the tree such that every
direct path from $\alpha'(0)$ always moves in the positive
direction. Consider
a single branching node $\alpha \in \Gamma $ and label the set of
segments radiating from it by $\branch_{\alpha}$. Let $\neb_{\alpha}$ denote the set of $z-1$ line segments
$k\in\branch_{\alpha}$ that radiate from $\alpha $ in a
positive direction, see Fig. \ref{fig:notation}(b).
Using these various
definitions we can introduce the idea of a generation. Take $\alpha_0$
to be the zeroth generation. The first generation then consists of the
set of nodes $\Sigma_1=\{\alpha(k),k\in \neb_{\alpha_0}\}$, the second
generation is $\Sigma_2=\{\alpha(l),l\in \neb_{\alpha},\alpha\in
\Sigma_1\}$ etc.

In the case of a homogeneous Cayley tree, each branch has the same length $L$, diffusion coefficient $D$ and velocity $v$. Let $x$, $0 \leq x \leq L$, be the position coordinate
along the $i$-th line segment with $0\leq i< \infty$. Using the given labeling of nodes, it follows that $x(\alpha'(i))=0$ and $x(\alpha(i))=L$. Let $p_i(x,t)$ denote
the probability density of a single particle to be at position $x$ on the $i$-th segment at time $t$. The density evolves according
to the Fokker-Planck equation
\begin{eqnarray}
\label{FPb}
  \frac{\partial p_i}{\partial t} = D\frac{\partial^2 p_i}{\partial
    x^2}-v\frac{\partial p_i}{\partial x} ,
  \quad 0 < x < L.
\end{eqnarray} 
As a further simplification, we assume that the initial position of the particle is on the primary branch,
\begin{equation}
p_i(x,0|x_0)= \delta(x-x_0)\delta_{i,0},\quad 0 < x_0<L.
\end{equation}
This initial condition means that all branches of a given generation are equivalent. Such a symmetry would also hold for the more general initial condition
\begin{equation}
p_i(x,0|x_0)= 2^{-n}\delta(x-x_0)\sum_{j,\alpha(j)\in \Sigma_n}\delta_{i,j},\quad 0 < x_0<L,
\end{equation}
which places the initial condition on any of the $2^n$ branches feeding into the nodes of the $n$-th generation with equal probability

Let $\flx[p_i]$ denote the corresponding probability current or flux, which is taken to be positive in the direction flowing away from the primary node at the soma:
\begin{equation}
  \flx[p]\equiv  - D\frac{\partial p}
        {\partial x} +v p.
\end{equation}
At all branch nodes $\alpha\in \Sigma_n$ of the $n$-th generation we impose the
continuity conditions
\begin{equation}
\label{cont}
  p_i(x(\alpha),t|x_0) = \Phi_{n}(x_0,t) ,\quad \mbox{for all}\ i \in {\mathcal I}_{\alpha},\ \alpha\in \Sigma_n,
\end{equation}
where the $\Phi_{n}(x_0,t)$ are unknown functions, which will ultimately be determined by imposing current conservation at each branch node:
\begin{equation}
\label{cons}
   \sum_{i \in{\mathcal I}_{\alpha}} \flx[p_i(x(\alpha),t|x_0)] = 0.
\end{equation}
Note that for the upstream segment $j\notin \overline{\mathcal I}_{\alpha}$, $x(\alpha)=L$ and the corresponding flux $\flx[p_j(L,t|x_0)]$ flows into the branch node, whereas for the remaining $z-1$ downstream segments $k\in 
\overline{\mathcal I}_{\alpha}$ we have $x(\alpha)=0$ and the flux $\flx[p_k(0,t|x_0)]$ flows out of the branch node. Finally, we impose either an absorbing or reflecting boundary condition on the primary branch at $x=0$:
\begin{equation}
p_0(0,t|x_0) =0 \quad \mbox{or}\quad  \flx[p_0](0,t|x_0)=0.
\end{equation}
In the latter case, we have the conservation condition
\begin{equation}
\label{normt}
\sum_{i\in \Gamma}\int_0^Lp_i(x,t|x_0)dx=1.
\end{equation}

\subsection{Steady-state solution (closed terminal node)} First, suppose that the terminal node is closed. In steady- state the current vanishes on each segment,
$\flx[p_i]\equiv 0$, so that the solution for any $i$ such that $\alpha(i)\in \Sigma_n$ is of the form
\begin{eqnarray}
p_i(x) =A_n\mbox{e}^{v x/D}.
\end{eqnarray}
Note that
\begin{equation}
A_n=\lim_{s\rightarrow 0}s\Phi_{n-1}(s,x_0),
\end{equation}
assuming the limit exists.
The continuity conditions (\ref{cont}) imply that the amplitudes
$A_n$ satisfy the iterative equation
\begin{eqnarray}
A_{n+1}=  A_n\mbox{e}^{vL/D}  \ \mbox{for all}\ n\geq 0,
\end{eqnarray}
with $A_0$ the steady-state solution at the origin.
Thus the amplitude $A_n$ at the $n$-th generation may be expressed in terms of the steady--state
concentration at the origin according to the relation 
\begin{eqnarray}
A_n=\mbox{e}^{nv L/D} A_0.
\end{eqnarray}
Imposing the normalization condition
\begin{eqnarray}
\sum_{i \in \Gamma}\int_0^L p_i(x)dx=1,
\end{eqnarray}
then yields the following equation for $A_0$,
\begin{eqnarray}
\label{A}
A_0^{-1}&=\frac{D}{v}\left [\e^{vL/D}-1\right ] \sum_{n=0}^{\infty} (z-1)^n\e^{nLv/D} .
\end{eqnarray}
This expresses $A_0^{-1}$ in terms of an infinite geometric series.
If this series is convergent then $A_0$ has a finite value and the steady--state is localized. On the other
hand, if the series diverges then $A_0 = 0$ and the steady--state is delocalized. The critical point of the LD transition is determined by the condition
\begin{equation}
(z-1)\e^{LvD}=1,
\end{equation}
which yields the critical velocity
\begin{equation}
\label{vc}
v_c=-\frac{D}{L}\ln(z-1) <0.
\end{equation}
Hence $v_c=0$ for $z=2$ (a semi-infinite line), $v_c\approx -0.69D/L$ for $z=3$ and $v_c\approx -1.1 D/L$ for $z=4$.
If $v<v_c$ then
\begin{eqnarray}
\label{A0}
A_0^{-1}
&=\frac{D}{|v| } \frac{1-\e^{-|v|L/D}}{1-(z-1)\e^{-L|v|/D}}.
\end{eqnarray}
In addition, the asymptotic decay of
the delocalized state exhibits conventional exponential behavior whereas at the critical point $v=v_c$ there is anomalous
behavior in the form of a critical slowing down \cite{Bressloff97}. Note that the point of criticality is determined by the dimensionless quantity
\begin{equation}
\mbox{Pe}\equiv \frac{Lv}{D},
\end{equation}
which is a P\'{e}clet number for the drift-diffusion process.
The critical P\'{e}clet number is then
\begin{equation}
\mbox{Pe}=-\ln(z-1).
\end{equation}

\subsection{Solution in Laplace space}

We now find the solution of equation (\ref{FPb}) in Laplace space for both the closed and open terminal node. Both types of solution will be needed when incorporating stochastic resetting into the model. We recently used an iterative method introduced in Ref. \cite{Newby09} to solve the open boundary problem. Here we extend the analysis to include the closed boundary value problem. First, Laplace transforming equations (\ref{FPb}), yields the following system of equations for any $i$ such that $\alpha(i)\in \Sigma_n$:
\begin{equation}
\label{AV}
\left [D\frac{\partial^{2} }{\partial x^{2}} 
     -v\frac{\partial}{\partial x} -s \right ]\tu_{i}(x,s) = -\delta_{i,0}\delta(x-x_0),
\end{equation}
together with the implicit boundary conditions
\begin{equation}
\tu_i(0,s) = \widetilde{\Phi}_{n-1}(s),\quad  \tu_i(L,s) = \widetilde{\Phi}_{n}(s).
\end{equation}
(For notational convenience, we drop the explicit dependence on the initial position $x_0$ and the velocity $v$ unless stated otherwise. We use $s$ as the Laplace variable in the absence of resetting, and the resetting rate $r$ as the Laplace variable when resetting is included, see Sect. 3-5.) If the terminal node is open then $\widetilde{\Phi}_{-1}(x_0,s)=0$ and 
$\widetilde{J}_0(s)>0$, whereas if the terminal node is closed then $\widetilde{\Phi}_{-1}(s)>0$ and 
$\widetilde{J}_0(s)=0$. The solution in each branch is given by the corresponding
finite interval Green's function ${\mathcal G}_L$ with homogeneous boundary conditions, together with terms satisfying the boundary conditions. That is,
  \begin{eqnarray}
  \label{GB}
   \tu_i(x,s) &=  \delta_{i,0}{\mathcal G}_L(x,x_0;s) +
    \widetilde{\Phi}_{n-1}(s)\widehat{F}(x,s)  + \widetilde{\Phi}_{n}(s)F(x,s)
  \end{eqnarray}
for $\alpha(i)\in \Sigma_n$.
The Green's function ${\mathcal G}_L$ satisfies
\begin{equation}
\label{AV2}
\left [D\frac{\partial^{2} }{\partial x^{2}} 
     -v\frac{\partial}{\partial x} -s \right ]{\mathcal G}_L(x,x_0;s) = -\delta(x-x_0),
\end{equation}
with ${\mathcal G}_L(0,x_0;s)=0={\mathcal G}_L(L,x_0;s)$.
The Green's function takes the form
\numparts
\begin{equation}
  \label{G}
 {\mathcal G}_L(x,x_0;s) = \left \{
  \begin{array}{cc}
    \frac{\displaystyle \uoo(x,s)\uoo(x_0-L,s)}{\displaystyle DW_L(s) },&0\leq x \leq x_0 \\ \\
    \frac{\displaystyle \uoo(x-L,s)\uoo(x_0,s)}{\displaystyle  DW_L(s) }, &x_0\leq x \leq L
  \end{array} \right . ,
\end{equation}
where
\begin{eqnarray}
\label{bag}
 \uoo(x,s) &= \e^{\mu_+(s)x} - \e^{\mu_-(s) x},
  \end{eqnarray}
  \begin{equation}
  \mu_{\pm}(s)=\frac{v}{2D}\pm \eta(s),\quad \eta(s)=\frac{\sqrt{v^2+4Ds}}{2D},
  \end{equation}
  and $W$ is the Wronskian
  \begin{eqnarray}
  W_L(s)  &= \uoo'(x_0,s)\uoo(x_0-L,s)-\uoo(x_0,s)\uoo'(x_0-L,s).
\end{eqnarray}
\endnumparts
The functions
$F(x,s)$ and $\widehat{F}(x,s)$ satisfy the homogeneous version of equation (\ref{AV}) with boundary conditions $F(0,s)=0,F(L,s)=1$ and $ \widehat{F}(0,s)=1,\widehat{F}(L,s)=0$:
\begin{eqnarray}
\label{Fk}
  F(x,s)=\frac{\psi(x,s)}{\psi(L,s)},\quad   \widehat{F}(x,s)=\frac{\psi(x-L,s)}{\psi(-L,s)}.
\end{eqnarray}

The unknown functions $\widetilde{\Phi}_{n}$ are now determined by imposing the
current conservation condition (\ref{cons}) at each branch node and using the identity
$\flx[\widetilde{\Phi}_{n} F]=\widetilde{\Phi}_{n} \flx[F]$, which follows from the observation that $\Phi_{n}$ is $x$--independent. 
At the zeroth generation node
${\alpha_{0}}$, the current conservation equation is given by (suppressing the $s$ variable)
  \begin{eqnarray}
\fl  \widetilde{\Phi}_{0} \flx[F](L)+\widetilde{\Phi}_{-1} \flx[\widehat{F}](L) = (z-1)
    \widetilde{\Phi}_{1}\flx[F](0) +(z-1)\widetilde{\Phi}_0\flx [\widehat{F}](0) + \targT,    \end{eqnarray}
  where 
\begin{equation}
 \targT(s)\equiv - \flx[{\mathcal G}_L](L)=\frac{\displaystyle \uoo'(0,s)\uoo(x_0,s)}{\displaystyle  W_L(s) },
\end{equation}
and at all  branching nodes $\alpha \in\Sigma_{n}$, $1\leq n$ we
have
\begin{eqnarray}
    \fl \widetilde{\Phi}_{n-1}\flx[\widehat{F}](L)
    +\widetilde{\Phi}_{n}\flx[F](L) 
    =(z-1)\widetilde{\Phi}_{n+1}\flx[F](0) +(z-1) \widetilde{\Phi}_{n}
    \flx[\widehat{F}](0).
  \end{eqnarray}
Note that $\targT$ depends on the source location $x_0$; this then generates an $x_0$--dependence of the functions $\widetilde{\Phi}_{n}$.

The four contributions to the probability flux at any branch $k\neq 0$ are
\numparts
\begin{eqnarray}
\label{gs}
 g(s)&\equiv \flx[\widehat{F}](L) =
  \frac{D\eta(s)\e^{{vL}/{2D}}}
  {\sinh(\eta(s) L)},\\
  h(s)&\equiv\flx[F](L)=
  -D\eta(s)\coth(\eta(s) L)
  +\frac{v}{2},\\
 \bar{g}(s)&\equiv \flx[F](0) =
 - \frac{D\eta(s) \e^{{-vL}/{2D}}}
  {\sinh(\eta(s) L)},\\
  \bar{h}(s)&\equiv\flx[\widehat{F}](0)=
  D\eta(s)\coth(\eta(s) L) +\frac{v}{2}.
  \label{hs}
\end{eqnarray}
\endnumparts
Note that the above flux functions are related according to the identities
\numparts
\label{ids}
\begin{eqnarray}
\fl  h(s)+\bar{h}(s)=v,\quad h(s)=\frac{v-g(s)+\bar{g}(s)}{2},\quad  \bar{h}(s)=\frac{v+g(s)-\bar{g}(s)}{2}.
\end{eqnarray}
\endnumparts
Using these definitions the current conservation equations simplify to
\begin{equation}
\label{pip}
g\widetilde{\Phi}_{-1}-H\widetilde{\Phi}_{0}
  +G\widetilde{\Phi}_{1} =  \targT
\end{equation}
on the primary branch, and
\begin{equation}
\label{itp}
  g\widetilde{\Phi}_{n-1}
  -H\widetilde{\Phi}_{n}
 +G\widetilde{\Phi}_{n+1}
  =   0
\end{equation}
for $n\geq 1$,
where
\begin{equation}
\label{H}
H= (z-1)\bar{h}-h,\quad G=-\bar{g}(z-1).
\end{equation}
The second-order difference equation (\ref{itp}) can be solved using the ansatz $\widetilde{\Phi}_n=\lambda^{n}\widetilde{\Phi}_0$, $n\geq 0$, which yields a quadratic equation for $\lambda$:
 \begin{equation}
 \label{quad}
 G\lambda^2-H\lambda+g=0.
 \end{equation}
 The two roots are
 \begin{equation}
 \label{lam0}
 \lambda_{\pm}=\frac{1}{2G}\left [H\pm \sqrt{H^2-4gG}\right ].
 \end{equation}
 It can be checked that $\lambda_{\pm}$ are real and $\lambda_{\pm } \leq 1$. However, we also find that $\lambda_+ \geq (z-1)^{-1}$ for all $s\geq 0$, which means that the total probability is not normalizable. In order to establish the latter result, consider
 the normalization condition
\begin{equation}
\label{norms}
{\mathcal N}\equiv \sum_{i\in \Gamma}\int_0^L\widetilde{p}_i(x,s)dx\leq \frac{1}{s},
\end{equation}
with the equality holding for a closed terminal node. 
Plugging in the solution (\ref{GB}) for $\widetilde{p}_i(x,s)$ yields the infinite series
\begin{eqnarray}
{\mathcal N}&=  \int_0^L {\mathcal G}_L(x,x_0;s) dx+ \widetilde{\Phi}_{-1}(s)\widehat{K}(s)  + \widetilde{\Phi}_{0}(s)K(s)\nonumber \\
&\quad +\sum_{n\geq 1} (z-1)^n\left [\widetilde{\Phi}_{n-1}(s)\widehat{K}(s)  + \widetilde{\Phi}_{n}(s)K(s)\right ],
\end{eqnarray}
with
\begin{eqnarray}
 K(s)&\equiv\int_0^L{F}(x,s)dx=-\frac{1}{s}[h(s)+g(s)],\\ \widehat{K}(s)&\equiv \int_0^L\widehat{F}(x,s)dx=\frac{1}{s}[\bar{h}(s)+\bar{g}(s)].
\end{eqnarray}
Taking $ \widetilde{\Phi}_{n}(s)=\lambda^n  \widetilde{\Phi}_{0}(s)$ leads to the infinite geometric series $\sum_{n\geq 0}(z-1)^n\lambda^n$, which is clearly divergent if $\lambda \geq 1/(z-1)$. Therefore, we set $\lambda=\lambda_-$ in the following

Substituting the solution for $\widetilde{\Phi}_1$ into equation (\ref{pip}) then implies that
\begin{equation}
 \label{phi0}
g\widetilde{\Phi}_{-1}=(H-G\lambda)\widetilde{\Phi}_{0}+\chi .
\end{equation}
Given $\widetilde{\Phi}_0$ and $\widetilde{\Phi}_{-1}$, the Laplace transform of the flux into the terminal node takes the explicit form (after reincorporating the dependence on the initial position)
\begin{eqnarray}
\label{Jtree}
\widetilde{J}(x_0,s)&=D\frac{\partial \widetilde{p}_0}{\partial x}(0,s|x_0)-v\widetilde{p}_0(0,s|x_0)\\
&=\left . D\frac{\partial {\mathcal G}_L(x,x_0;s)}{\partial x} \right |_{x=0} +D\left . \widetilde{\Phi}_{0}(x_0,s)\frac{\partial F(x,s)}{\partial x}\right |_{x=0}\nonumber \\
&\quad +D\left . \widetilde{\Phi}_{-1}(x_0,s)\frac{\partial \widehat{F}(x,s)}{\partial x}\right |_{x=0}-v\widetilde{\Phi}_{-1}(x_0,s).\nonumber 
\end{eqnarray}
In the case of an open terminal node, we have $\widetilde{\Phi}_{-1}=0$ in  equation (\ref{phi0}) so that
\begin{equation}
\widetilde{\Phi}_{0}=\frac{\chi}{G\lambda -H},
\end{equation}
and the flux through the node is
\begin{eqnarray}
\label{Jopen}
\fl \widetilde{J}(x_0,s)&=\left . D\frac{\partial {\mathcal G}_L(x,x_0;s)}{\partial x} \right |_{x=0}  -\frac{D\chi(s)}{H(s)-\lambda(s) G(s)}\left . \widetilde{\Phi}_{0}(x_0,s)\frac{\partial F(x,s)}{\partial x}\right |_{x=0}.
\end{eqnarray}
On the other hand, when the terminal node is closed, the unknown functions $\widetilde{\Phi}_{-1},\widetilde{\Phi}_{0}$ are found by solving the pair of equations (\ref{phi0}) and (\ref{Jtree}) with $\widetilde{J}(x_0,s)=0$.

\subsection{Semi-infinite line ($z=2$)} If the coordination number is $z=2$, then the tree reduces to a semi-infinite line, which is much easier to analyze. 
In particular, for a closed terminal node the solution of equation (\ref{AV}) is $\widetilde{p}_i(x,s)=\widehat{\mathcal G}(x+nL,s)$ for $\alpha(i)\in \Sigma_n$,
where $\widehat{\mathcal G}$ is
the Greens function \begin{equation}
\label{1dAV2}
\left [D\frac{\partial^{2} }{\partial x^{2}} 
     -v\frac{\partial}{\partial x} -s \right ]\widehat{\mathcal G}(x,x_0;s) = -\delta(x-x_0),
\end{equation}
with 
\begin{equation}
D\partial_x\widehat{\mathcal G}(0,x_0;s)-v\widehat{\mathcal G}(0,x_0;s)=0,\quad \widehat{\mathcal G}(x,x_0;s)\rightarrow 0 \mbox{ as } x\rightarrow \infty.
\end{equation}
The Green's function $\widehat{\mathcal G}$ has the explicit form
\numparts
\begin{equation}
  \label{1dGhat}
\widehat{\mathcal G}(x,x_0;s) = \left \{
  \begin{array}{cc}
    \frac{\displaystyle \phi_1(x,s)\phi_2(x_0,s)}{\displaystyle D\widehat{W}(s) },&0\leq x \leq x_0 \\ \\
    \frac{\displaystyle \phi_1(x_0,s)\phi_2(x,s)}{\displaystyle  D\widehat{W}(s) }, &x_0\leq x 
  \end{array} \right . ,
\end{equation}
where
\begin{eqnarray}
\label{1dbag}
  \phi_1(x,s) &= \left [(\mu_-(s)-v/D)\e^{\mu_+(s)x} -(\mu_+(s)-v/D) \e^{\mu_-(s)x}\right ],\\ \phi_2(x,s)&= \e^{\mu_-(s)x},
  \end{eqnarray}
and $\widehat{W}$ is the Wronskian
  \begin{eqnarray}
  \widehat{W}(s)  &= \phi_1'(x_0,s)\phi_2(x_0,s)-\phi_1(x_0,s)\phi_2'(x_0,s).
\end{eqnarray}
\endnumparts
It follows that at the terminal node,
\begin{equation}
\label{Phi1d}
\widetilde{\Phi}_{-1}(x_0,s) =\widehat{\mathcal G}(0,x_0;s)= \frac{\displaystyle \e^{-\mu_+(s)x_0}}{\mu_+(s) }.
\end{equation}
It is clear that the branch length $L$ plays no role when $z=2$, unless there is quenched disorder, see Sect. 5.

When the terminal node is open we simply replace the function $\phi_1$ by $\psi$ to obtain the Green's function
\numparts
\begin{equation}
  \label{1dG}
{\mathcal G}(x,x_0;s) = \left \{
  \begin{array}{cc}
    \frac{\displaystyle\psi(x,s)\phi_2(x_0,s)}{\displaystyle D{W}(s) },&0\leq x \leq x_0 \\ \\
    \frac{\displaystyle \psi(x_0,s)\phi_2(x,s)}{\displaystyle  D{W}(s) }, &x_0\leq x 
  \end{array} \right . ,
\end{equation}
where
  \begin{eqnarray}
  {W}(s)  &= \psi'(x_0,s)\phi_2(x_0,s)-\psi(x_0,s)\phi_2'(x_0,s).
\end{eqnarray}
\endnumparts
The flux through the open terminal node is
\begin{equation}
\label{J1d}
\widetilde{J}(x_0,s)=D\partial_x\widehat{\mathcal G}(0,x_0;s)=\e^{-\mu_+(s)x_0}.
\end{equation}
It can be checked numerically that the simpler expressions (\ref{Phi1d}) and (\ref{J1d}) are equivalent to the corresponding results obtained by taking $z=2$ in the general solution for a Cayley tree.

\setcounter{equation}{0}

\section{Steady-state solution with resetting (closed terminal node)}

Now suppose that the particle can instantaneously reset to its initial position on the primary branch according to a Poisson process with rate $r$. Denoting the probability density with reset by $P_i$ on the $i$-th branch, equations (\ref{FPb}) are modified according to
\begin{eqnarray}
  \frac{\partial P_i}{\partial t} = D\frac{\partial^2 P_i}{\partial
    x^2}-v\frac{\partial P_i}{\partial x} -rP_i(x,t) +r\delta(x-x_0)\delta_{i,0}
   \label{FPr}
  \end{eqnarray} 
  for $0 < x <L$ and $i\in \Gamma$. 
 In order to calculate the steady-state solution at the closed terminal node,  it is convenient to work with a renewal equation approach \cite{Evans20}. In the absence of resetting ($r=0$), equation (\ref{FPr}) reduces to equation (\ref{FPb}) whose solution is given by (\ref{GB}) in Laplace space. When resetting is included, the probability density $P_i(x,t|x_0)$ has two distinct types of contribution: paths where no resetting events have occurred up to time $t$, and paths where the last resetting event occurred at time $\tau_l=t-\tau$ for some $\tau \in (0,t)$. The probability density of no resetting events up to time $t$ is $\e^{-rt}$. Similarly, the probability density that the last resetting event occurred at time $t-\tau$ (with no subsequent resetting events) is $r\e^{-r\tau}$. Using the fact that in the latter case $X(t-\tau)=x_0$ and one has pure drift-diffusion over the time interval $(t-\tau,t)$, the full time-dependent solution to equation (\ref{FPr}) satisfies the so-called last renewal equation
\begin{equation}
\label{GBr}
P_i(x,t|x_0)=\e^{-rt}p_i(x,t|x_0)+r\int_0^tp_i(x,\tau|x_r)\e^{-r\tau}d\tau.
\end{equation}

The steady-state solution with resetting, $P_i^*(x,x_0)$, is obtained by taking the limit $t\rightarrow \infty$ in equation (\ref{GBr}):
\begin{equation}
\label{snode}
P_i^*(x,x_0)=r\int_0^{\infty}p_i(x,\tau|x_0)\e^{-r\tau} d\tau=r\widetilde{p}_i(x,r|x_0).
\end{equation}
That is, $P_i^*(x,x_0)$ is determined by the $r$-Laplace transform of the solution without resetting, see equation (\ref{GB}). Consider, in particular, the effect of resetting on the steady-state solution at the terminal node, which is given by
\begin{equation}
\Phi^*(x_0,r)\equiv r\widetilde{p}_0(0,r|x_0) =r\widetilde{\Phi}_{-1}(x_0,r),
\end{equation}
with $\widetilde{\Phi}_{-1}(x_0,r)$ obtained from equations (\ref{phi0}) and (\ref{Jtree}) after setting $s=r$ and $\widetilde{J}(x_0,s)=0$:
\begin{eqnarray}
\label{Phi1}
\widetilde{\Phi}_{-1}(x_0,r) =\frac{C_a(x_0,r)}{C_b(x_0,r)},
\end{eqnarray}
where
\begin{eqnarray*}
\fl C_a(x_0,r)&= -\frac{\displaystyle \uoo'(0,s)\uoo(x_0-L,s)}{\displaystyle  W(s) }[H-\lambda G]-D\frac{\psi'(0,s)}{\psi(L,s)}\frac{\displaystyle \uoo'(0,s)\uoo(x_0,s)}{\displaystyle  W(s) },\\
\fl C_b(x_0,r)&=gD\frac{\psi'(0,s)}{\psi(L,s)}+\left (D\frac{\psi'(-L,s)}{\psi(-L,s)}-v\right )[H-\lambda G].
\end{eqnarray*}
When $z=2$ we can use the much simpler expression (\ref{Phi1d}),
\begin{equation}
\label{Phi1d2}
\widetilde{\Phi}_{-1}(x_0,r) = \frac{\displaystyle \e^{-\mu_+(r)x_0}}{\mu_+(r) }.
\end{equation}

\begin{figure}[t!]
\raggedleft
\includegraphics[width=13cm]{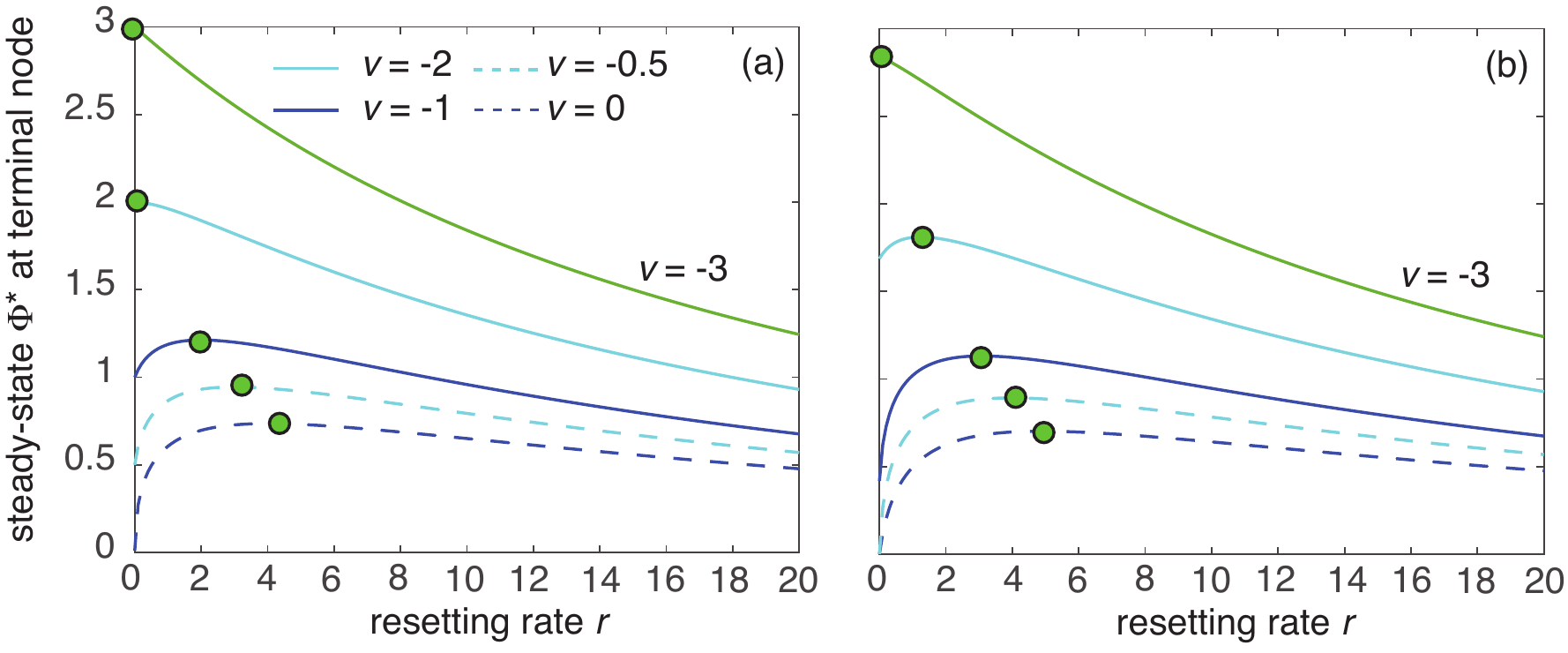} 
\caption{Drift-diffusion with resetting. Plot of steady-state $\Phi^*(x_0,r)$ at a closed terminal node as a function of the resetting rate for different velocities $v$ and (a) $z=2$, (b) $z=3$. Other parameters are $x_0=0.5$, $D=1$ and $L=1$. Filled circles indicate maxima. }
\label{fig2}
\end{figure}

\begin{figure}[b!]
\raggedleft
\includegraphics[width=13cm]{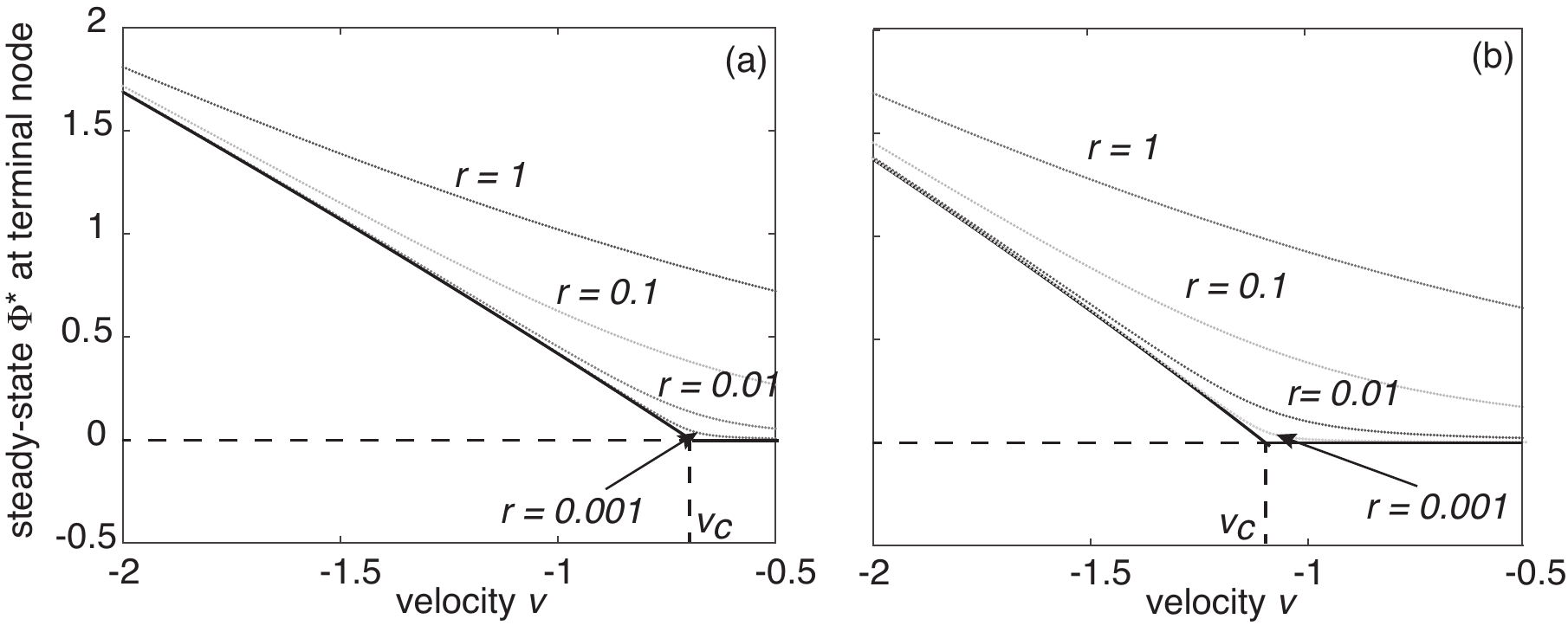} 
\caption{Drift-diffusion with resetting. Plot of steady-state $\Phi^*(x_0,r)$ as a function of the velocity $v$ for various values of the resetting rate $r$, with $\Phi_{-1}(x_0,r)$ the $r$-Laplace transformed solution at the terminal node. (a) $z=3$. (b) $z=4$. Other parameter values are $D=L=1$ and $x_0=0.5$. The solid black curve is the steady-state solution $A_0$ determined by equation (\ref{A0}). Note that $A_0=0$ for $v\geq v_c$, where $v_c$ is the critical velocity (\ref{vc}).}
\label{fig3}
\end{figure}

\begin{figure}[b!]
\raggedleft
\includegraphics[width=7cm]{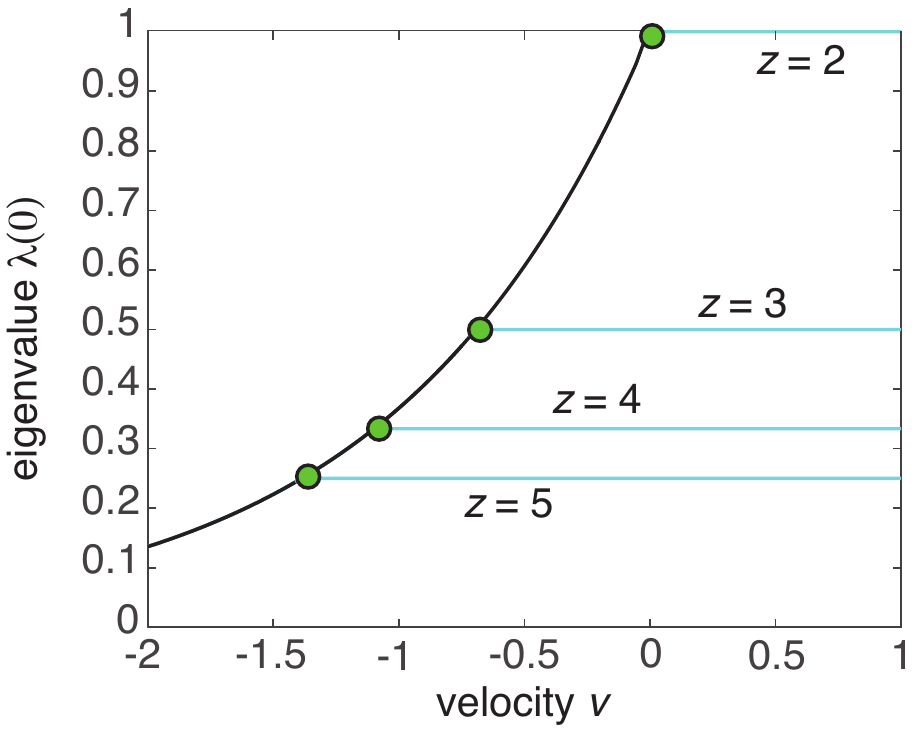} 
\caption{Plot of eigenvalue $\lambda(0)$ as a function of the velocity $v$ for different coordination numbers $z$. Fo each $z$, the horizontal line is $1/(z-1)$ which terminates at the critical velocity $v_c$ (indicated by the filled circle). For $v<v_c(z)$, $\lambda(0)= \e^{vL/D}$ as shown by the exponential curve. Other parameters are $x_0=0.5$, $D=1$ and $L=1$.}
\label{fig4}
\end{figure}

In Fig. \ref{fig2}, we plot $\Phi^*(x_0,r)$ as a function of the resetting rate $r$ for various velocities and initial positions $x_0$. Two particular results can be obtained from these figures. First, for $v>v_c$ we have $\Phi^*(x_0,r)\rightarrow 0$ as $r\rightarrow 0$. This reflects the LD transition in the absence of resetting discussed in Sect. 2.1. Second, for a range of velocities there is a unique peak in $\Phi^*(x_0,r)$ at a $v$-dependent resetting rate $r^{**}(v)>0$.
The first result is explored further in Fig. \ref{fig3}, where  we plot $\Phi^*(x_0,r)$ as a function of the velocity $v$ for various resetting rates $r$. It can be seen that in the limit $r\rightarrow 0$, the solution converges to the steady-state amplitude $A_0$ at the terminal node, equation (\ref{A0}). That is,
$A_0=\lim_{r\rightarrow 0}r\widetilde{\Phi}_{-1}(x_0,r )$. In particular,  $A_0=0$ for $v\geq v_c$ where $v_c$ is the critical velocity (\ref{vc}). The sharp transition at $r=0$ can be understood as follows. Setting $s=0$ in equations (\ref{gs})-(\ref{hs}) with $\eta(0)=v/2D$ yields
\begin{eqnarray}
\label{ids2}
&g(0)+\bar{g}(0)=v,\quad h(0)=\bar{g}(0),\quad  \bar{h}(0)= {g}(0).
\end{eqnarray}
Equation (\ref{H}) then implies that
\begin{eqnarray*}
&H^2(0)-4g(0)G(0)=((z-1)\bar{h}(0)-h(0))^2+4g(0)\bar{g}(0)(z-1)\\
&=((z-1)g(0)-\bar{g}(0))^2+4g(0)\bar{g}(0)(z-1)=((z-1)g(0)+\bar{g}(0))^2,
\end{eqnarray*}
and thus
\begin{equation}
\lambda(0)=-\frac{(z-1){g}(0)-\bar{g}(0)-|(z-1)g(0)+\bar{g}(0)|}{2\bar{g}(0)(z-1)}.
\end{equation}
If $(z-1){g}(0)+\bar{g}(0)>0$, then
\begin{equation}
\lambda(0)=\frac{1}{z-1}.
\end{equation}
On the other hand, if 
$(z-1){g}(0)+\bar{g}(0)<0$, then
\begin{equation}
\lambda(0)=-\frac{g(0)}{\bar{g}(0)}=\e^{vL/D}.
\end{equation}
The transition point is given by $v=v_c$. The behavior of $\lambda(0)$ as a function of $v$ is illustrated in Fig. \ref{fig4}. Now multiplying both sides of equation (\ref{phi0}) by $s$ and taking the limit $s\rightarrow 0$ yields
\numparts
\begin{eqnarray}
\label{boa}
A_0&= (z-1)A_1 +O(s) \mbox{ for } v>v_c,\\ A_0&= -\frac{g(0)}{\bar{g}(0)} A_1 +O(s) \mbox{ for } v<v_c.
\label{bob}
\end{eqnarray}
\endnumparts
Similarly, multiplying the right-hand side of equation (\ref{Jtree}) by $s$ and taking the limit $s\rightarrow 0$ gives the no-flux condition
\begin{eqnarray*}
0&=A_1F'(0,0)D +A_0 [\widehat{F}'(0,0)D-v]=-\bar{g}(0)A_1-\bar{h}(0)A_0.
\end{eqnarray*}
For $v>v_c$ we have
\[A_0[(z-1)^{-1}\bar{g}(0)+\bar{h}(0)]=0 \Rightarrow A_0=0,
\]
whereas for $v<v_c$,
\[A_0[g(0)-\bar{h}(0)]=0\cdot A_0 =0.
\]
In the latter case, one has to go to $O(s)$ to recover equation (\ref{A0}) for $A_0$.

\begin{figure}[b!]
\raggedleft
\includegraphics[width=7cm]{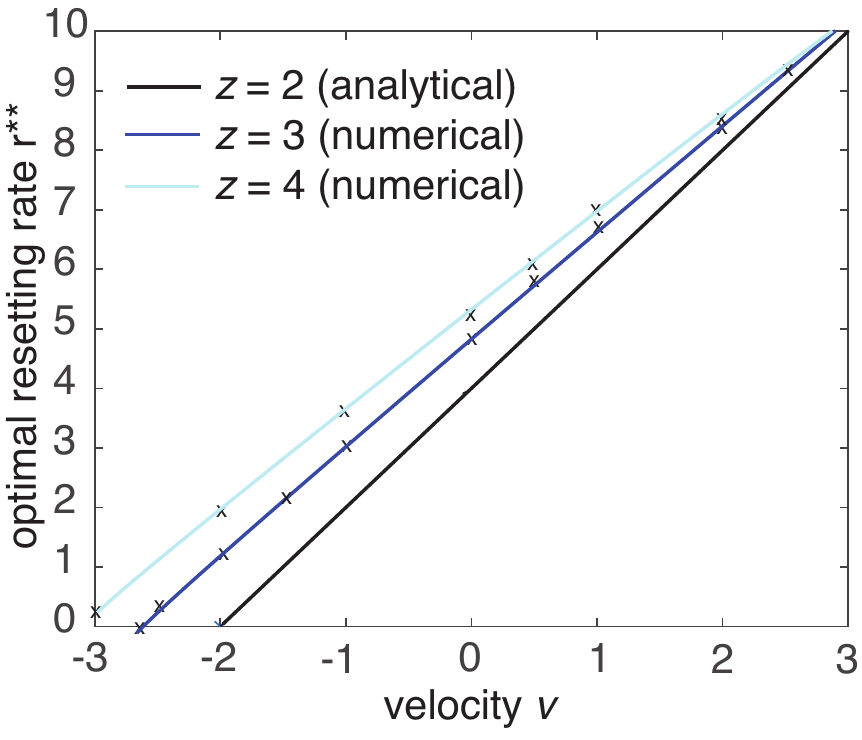} 
\caption{Plot of optimal resetting rate $r^{**}(v)$ at which $\Phi^*(x_0,r)$ has a maximum as a function of the velocity $v$ for $z=2,3,4$ Points of intersection with the $v$-axis determine the critical velocity $v_c^{**}(z)$. Other parameters are $x_0=0.5$, $D=1$ and $L=1$.}
\label{fig5}
\end{figure}

Turning to the second result, we find that there exists a new critical velocity $v_{c}^{**}$, $v_c^{**}< v_c$, such that the peak of $\Phi^*(x_0,r)$ occurs at a non-zero resetting rate $r^*(v)>0$ when $v>v_c^{**}$. On the other hand, $r^*(v)=0$ for $v<v_c^{**}$. In both cases, the particle tends to be localized around $x_0$ for sufficiently large $r$ and this reduces $\Phi^*(x_0,r)$. Increasing $r$ from zero can initially increase $\Phi^*(x_0,r)$ provided that the drift towards the terminal node is not too strong. The critical velocity $v_c^{**}$ is obtained from the condition
\begin{equation}
\left .\frac{d\left [r\widetilde{\Phi}_{-1}(x_0,r)\right ]}{dr}\right |_{r=r^*}=0.
\end{equation}
For simplicity, consider the semi-infinite line $z=2$. Substituting equation (\ref{Phi1d}) into the above equation yields 
\begin{equation}
\frac{d\mu_+}{dr}[\mu_+x_0+1]=\frac{\mu_+}{r}.
\end{equation}
Since 
\[\mu_+'(r)=\frac{1}{2D\eta(r)}=\frac{1}{2D\mu_+(r) -v},
\]
we find that $r$ satisfies the equation
\begin{equation*}
(rx_0-v)\sqrt{v^2+4Dr}= v^2+2Dr -rvx_0.
\end{equation*}
Squaring both sides and rearranging finally gives a cubic for $r$ of the form
\[r^2[rx_0^2-vx_0-D]=0.\]
One solution is $r=0$ and the other is 
\begin{equation}
\label{rstars}
r^{**}=\frac{D+vx_0}{x_0^2},
\end{equation}
provided that the right-hand side is positive. Hence, the critical velocity $v_c^{**}$ for $z=2$ is determined by the condition
\begin{equation}
\label{Pe0}
\mbox{Pe}'\equiv \frac{vx_0}{D}= -1,
\end{equation}
where $\mbox{Pe}'$ is a second P\'{e}clet number. That is, $v_c^{**}=-D/x_0$.
If the initial position $x_0=0.5$ and $D=1$, then $v_c'=-2$, which is consistent with Fig. \ref{fig2}(a). In Fig. \ref{fig5}
 we plot $r^{**}(v)$ as a function of $v$ for $z=2,3,4$. The straight line for $z=2$ is based on equation (\ref{rstars}), whereas the curves for $z=3,4$ are obtained by numerically solving equations (\ref{phi0}) and (\ref{Jtree}) with $\widetilde{J}(x_0,s)=0$. We find that the $z=3,4$ curves asymptotically approach the $z=2$ line as $v\rightarrow \infty$. Similar results hold for higher coordination numbers with $v_c^{**}(z)$ a decreasing function of $z$.

\setcounter{equation}{0}

\section{Hitting probability and MFPT with resetting (open terminal node)}

\begin{figure}[t!]
\raggedleft
\includegraphics[width=13cm]{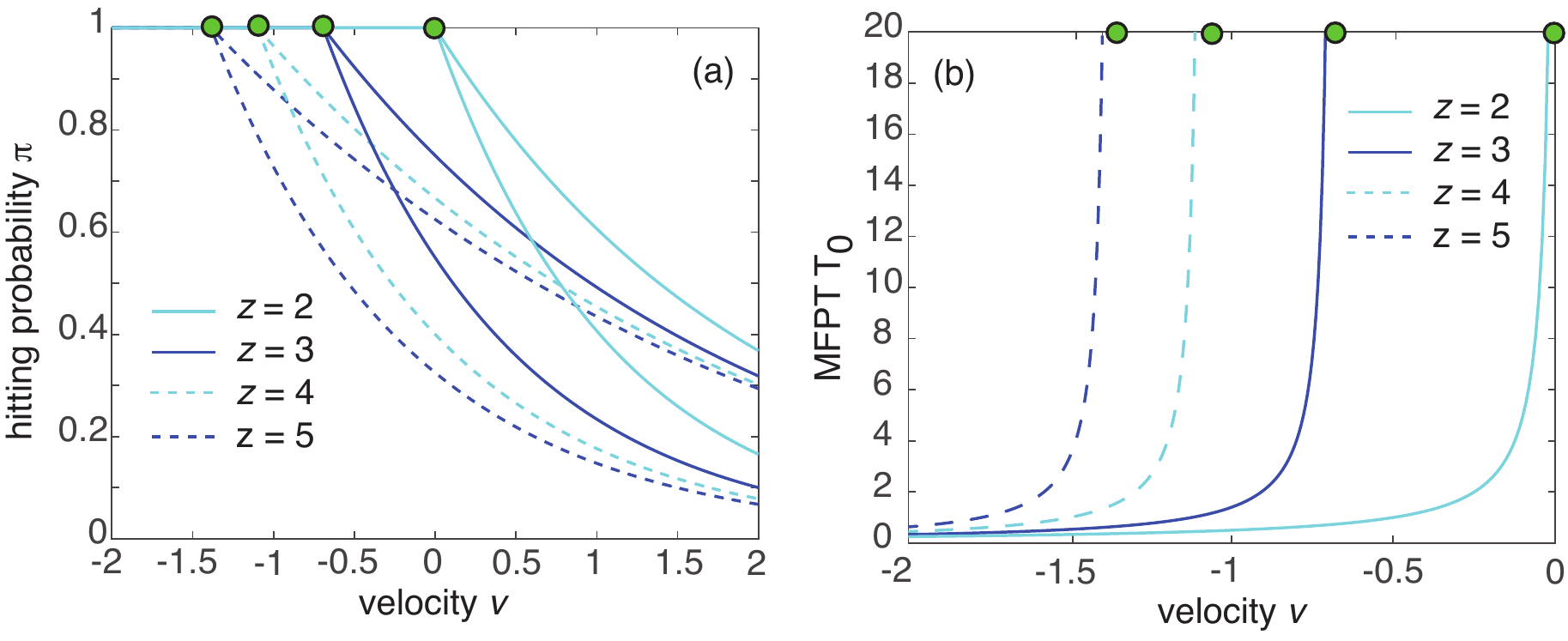} 
\caption{(a) Plot of hitting probability $\pi_0$ without resetting as a function of velocity $v$ for different coordination numbers $z$. For each $z$, the upper curve is for the initial position $x_0=0.5$ and the lower curve is for $x_0=0.9$. (b) Corresponding plot of MFPT $T_0$ without resetting for $x_0=0.5$. Other parameters are $D=1$ and $L=1$. The filled circles indicate the critical velocity $v_c(z)$ for each value of $z$.}
\label{fig6}
\end{figure}

Two quantities of interest in the case of an open terminal node are the hitting probability that a particle starting at $x_0$ on the primary branch of the tree is eventually absorbed by the primary node, and the corresponding MFPT. First consider these quantities in the absence of resetting. Let ${\mathcal T}_0(x_0)$ denote the FPT that the particle is captured by the terminal node with ${\mathcal T}_0(x_0)=\infty$ indicating that it is not captured. (The zero subscript indicates that there is no resetting.) 
Let $Q_0(x_0,t)$ be the survival probability that the particle remains free in the time interval $[0,t)$, given that it started at $x_0$:
\begin{eqnarray}
\label{q00}
Q_0(x_0,t)&=\P[t<{\mathcal T}_0(x_0)<\infty ]= \int_{\Gamma}p(x,t|x_0)dx.
\end{eqnarray}
Differentiating both sides with respect to $t$, using the FP equation (\ref{FPb}) and imposing current conservation shows that
\begin{equation}
\frac{\partial Q_0(x_0,t)}{\partial t} =-J(x_0,t),
\end{equation}
where $J(x_0,t)$ is the flux into the terminal node:
\begin{eqnarray}
J(x_0,t)&=D\frac{\partial {p}_0}{\partial x}(0,t|x_0)-vp_0(0,t|x_0)=D\frac{\partial {p}_0}{\partial x}(0,t|x_0),
\label{Jk}
\end{eqnarray}
after imposing the absorbing boundary condition.
In Laplace space,
\begin{eqnarray}
\label{qLT}
s\widetilde{Q}_0(x_0,s)-1&=- \widetilde{J}(x_0,s).
\end{eqnarray}
The hitting probability is given by
\begin{eqnarray}
\pi_0(x_0) & =1-\lim_{t\rightarrow \infty}Q_0(x_0,t)=1-\lim_{s\rightarrow 0}s\widetilde{Q}_0(x_0,s)= \widetilde{J}(x_0,0).
\label{pi}
\end{eqnarray}
The corresponding MFPT (if it exists) is defined according to
\begin{eqnarray}
\fl T_0(x_0)&\equiv \E[{\mathcal T}_0(x_0)]=- \int_0^{\infty}t \frac{dQ_0(x_0,t)}{dt}d\tau=\int_0^{\infty}Q_0(x_0,t)dt = \widetilde{Q}_0(x_0,0).
\end{eqnarray}
We have used the fact that the FPT density $f_0(x_0,t)$ is related to the survival probability according to $f_0=-dQ_0/dt$. Example plots of $\pi_0(x_0)$ as a function of the velocity $v$ are shown in Fig. \ref{fig6}(a). It can be seen that there is a critical phase transition from recurrent ($\pi_0 =1$) to transient ($\pi_0 < 1$) transport at the LD threshold $v_c$; for $v>v_c$, the hitting probability is a decreasing function of both $v$ and $z$, and the corresponding MFPT is infinite. The last point is illustrated in Fig. \ref{fig6}(b). The change in behavior of the hitting probability can be understood by setting $s=0$ in equations (\ref{phi0}) and (\ref{Jopen}) with $\widetilde{\Phi}_{-1}=0$ and $v<v_c$. In particular,
\begin{equation}
\widetilde{\Phi}_{0}(x_0,0)=\frac{\chi(0)}{\lambda(0) G(0)-H(0)}=\frac{\widehat{F}'(0,0)}{D F'(0,0)}\frac{\psi(x_0,0)}{\psi'(0,0)}
\end{equation}
and
\begin{equation}
\widetilde{J}(x_0,0) = -{F}'(0,0) \frac{\psi(x_0-L,0)}{\psi'(0,0)} +\widehat{F}'(0,0)\frac{\psi(x_0,0)}{\psi'(0,0)}=1.
\end{equation}

Now consider the effects of stochastic resetting. Let $Q_r(x_0,t)$ denote the survival probability for resetting at a rate $r$:
\begin{eqnarray}
\label{rq00}
Q_r(x_0,t)&=\P[t<{\mathcal T}_r(x_0)<\infty ]= \int_{\Gamma}P(x,t|x_0)dx,
\end{eqnarray}
where ${\mathcal T}_r(x_0)$ is the corresponding FPT. As in the analysis of the steady-state solution, $Q_r$ can be related to the survival probability without resetting, $Q_0$, using a last renewal equation \cite{Evans11a,Evans11b}:
\begin{eqnarray}
\label{renQ}
Q_r(x_0,t)=&\e^{-rt}Q_0(x_0,t)+r\int_0^tQ_0(x_0,\tau)Q_r(x_0,t-\tau)\e^{-r\tau}d\tau.  
\end{eqnarray}
The first term on the right-hand side represents trajectories with no resettings. The integrand in the second term is the contribution from trajectories that last reset at time $\tau\in (0,t)$, and consists of the product of the survival probability starting from $x_0$ with resetting up to time $t-\tau$ and the survival probability starting from $x_0$ without any resetting over the time interval $\tau$. Laplace transforming the last renewal equation, using the convolution theorem and rearranging shows that
\begin{equation}
\label{Qr}
{ \widetilde{Q}_r(x_0,s)=\frac{ \widetilde{Q}_0(x_0,r+s)}{1-r \widetilde{Q}_0(x_0,r+s)}.}
 \end{equation}
 It immediately follows that the hitting probability with resetting, $\pi_r(x_0)$, is unity for all $r>0$ and $v$. The corresponding MFPT is given by
\begin{eqnarray}
\fl T_r(x_0)&\equiv \E[{\mathcal T}_r(x_0)]=- \int_0^{\infty}t \frac{dQ_r(x_0,t)}{dt}d\tau=\int_0^{\infty}Q_r(x_0,t)dt = \widetilde{Q}_r(x_0,0).
\label{TQ}
\end{eqnarray}
From equation (\ref{TQ}), we have
\begin{equation}
T_r(x_0) =\widetilde{Q}_r(x_0,0)=\frac{ \widetilde{Q}_0(x_0,r)}{1-r \widetilde{Q}_0(x_0,r)}.
 \end{equation}
 Substituting for $\widetilde{Q}_0(x_0,r)$ using equation (\ref{qLT}) then yields
 \begin{equation}
 \label{Tr}
T_r(x_0) =\frac{ 1-\widetilde{J}(x_0,r)}{r\widetilde{J}(x_0,r)},
 \end{equation}
 where $\widetilde{J}(x_0,r)$ is the Laplace transformed flux into the terminal node without resetting, see equation (\ref{Jopen}). 
 
 \begin{figure}[b!]
\raggedleft
\includegraphics[width=13cm]{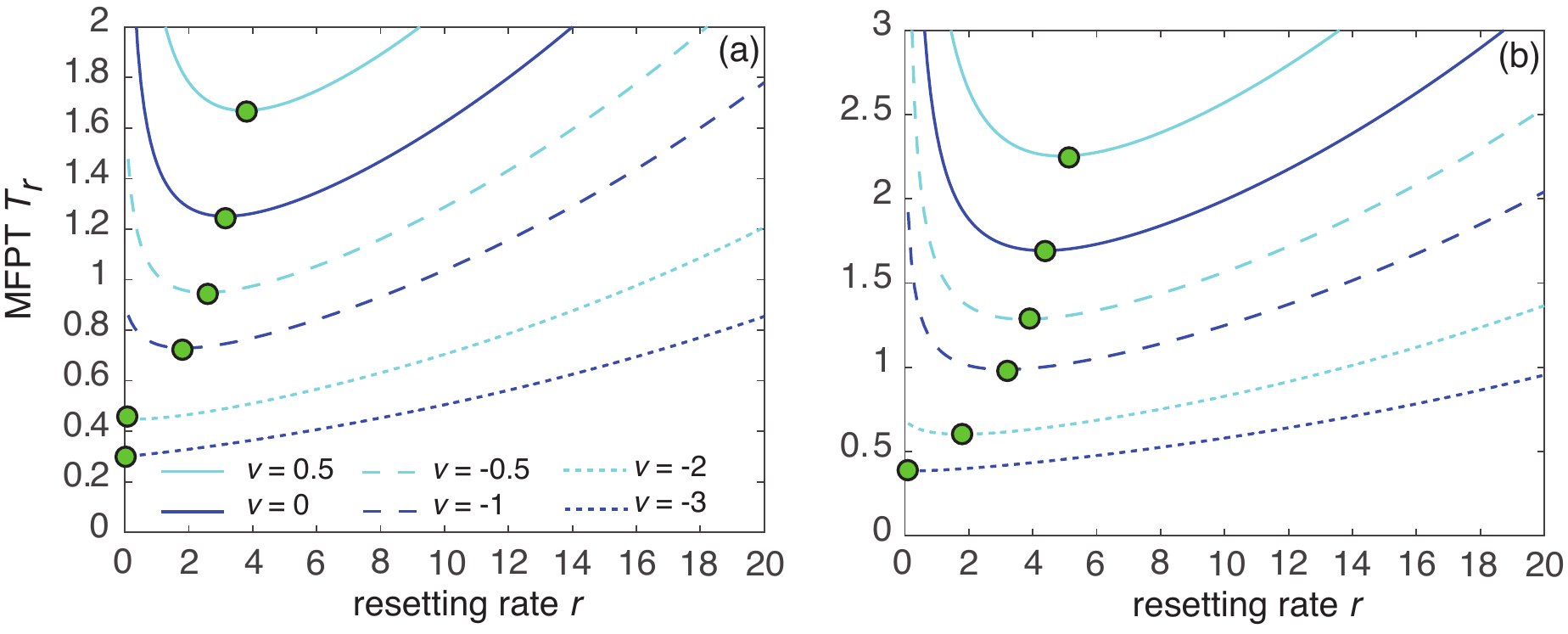} 
\caption{Drift-diffusion with resetting. (a) Plot of MFPT $T_r(x_0)$ to be absorbed at an open terminal node as a function of the resetting rate $r$ for different velocities $v$ and  (a) $z=2$ (b) $z=3$. Other parameters are $x_0=0.9$, $D=1$ and $L=1$. Filled circles indicate minima.}
\label{fig7}
\end{figure}

\begin{figure}[b!]
\raggedleft
\includegraphics[width=7cm]{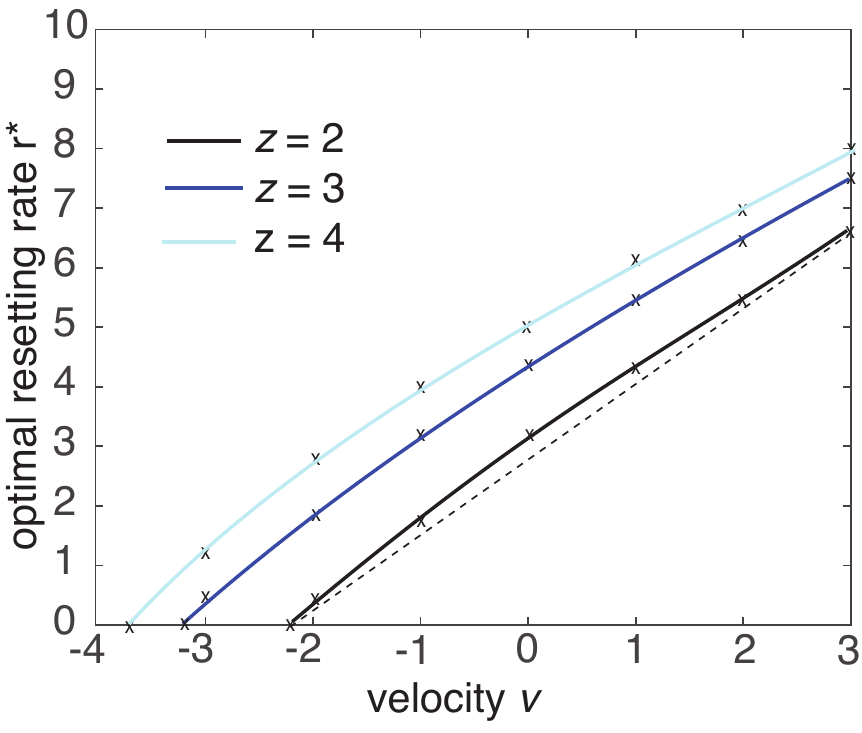} 
\caption{Plot of optimal resetting rate $r^{*}(v)$ at which $T_r(x_0)$ has a minimum as a function of the velocity $v$ for $z=2,3,4$ Points of intersection with the $v$-axis determine the critical velocity $v_c^{*}(z)$. Other parameters are $x_0=0.9$, $D=1$ and $L=1$. The dashed straight line highlights the fact that the graph of $r^*(v)$ is curved when $z=2$.}
\label{fig8}
\end{figure}

\begin{figure}[t!]
\raggedleft 
\includegraphics[width=7cm]{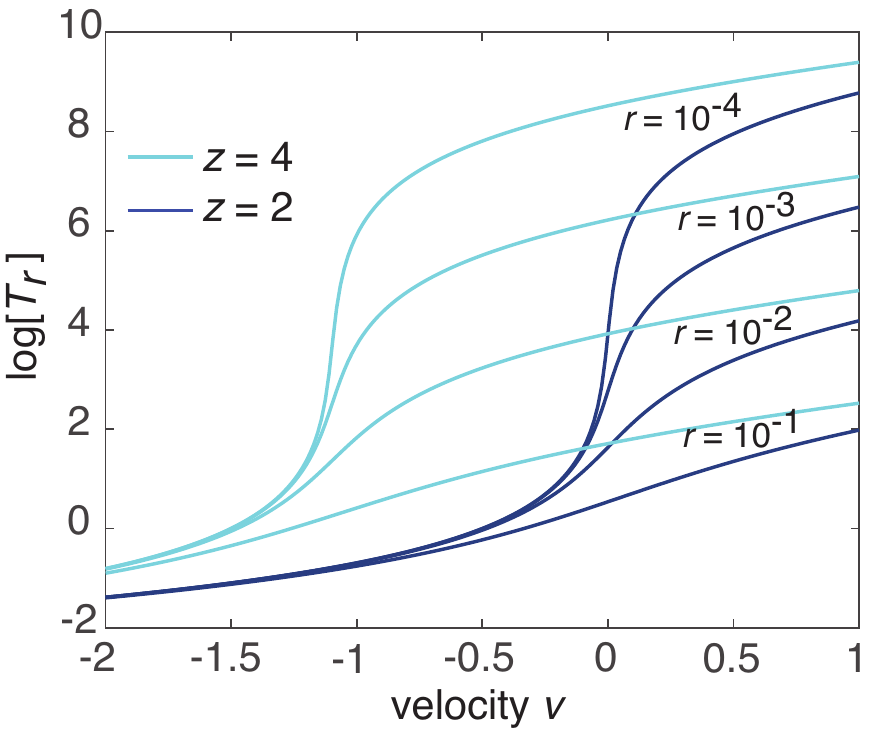} 
\caption{Drift-diffusion with resetting. Plot of MFPT $T_r(x_0)$ as a function of the velocity $v$ for decreasing values of $r$ and for $z=2,4$. Other parameters are $x_0=0.5$, $D=1$ and $L=1$. There is a rapid increase of the MFPT as $v$ crosses the critical velocity $v_c(z)$.}
\label{fig9}
\end{figure}

 In Fig. \ref{fig7} we plot the MFPT $T_r(x_0)$ as a function of $r$ for fixed $x_0$. Analogous to the steady-state at a closed terminal node, we find that for a range of velocities $v$, the MFPT $T_r(x_0)$ is a unimodal function of $r$ with a unique minimum at an optimal resetting rate $r^*(v)$. Moreover, there exists a critical velocity $v_c^*$ such that $r^*(v)=0$ for $v<v^*_c$ and $r^*(v)>0$ for $v>v^*_c$. This is a generalization of the phase transition identified in \cite{Ray19,Pal19} for the semi-infinite line. Indeed, we recover the analytical results of these papers for $z=2$. (The only difference is that the previous authors consider a target at $x=x_0$ and take the origin to be the reset point.) Substituting the solution (\ref{J1d}) into (\ref{Tr}) gives
 \begin{equation}
\label{J2d}
T_r(x_0) =\frac{ 1-\e^{-\mu_+(r)x_0}}{r\e^{-\mu_+(r)x_0}} =\frac{ \e^{\mu_+(r)x_0}-1}{r} .
\end{equation}
The optimal resetting rate is obtained from the condition
\begin{eqnarray}
0=\frac{dT_r}{dr}=-\frac{ \e^{\mu_+(r)x_0}-1}{r^2} +\frac{\mu_+'(r)x_0\e^{\mu_+(r)x_0}}{r}.
 \end{eqnarray}
 This yields a transcendental equation for $\mu_+(r)$ of the form
 \begin{equation}
 \frac{rx_0^2/D}{1-  \e^{-\mu_+(r)x_0}}=2x_0\mu_+(r)-\mbox{Pe}'.
 \end{equation}
 Using a graphical construction it can be shown that there exists a positive solution for $r=r^*$ provided that $\mbox{Pe}' >-2$ \cite{Ray19}, that is, $v>v_c^* =-2D/x_0=2v_c^{**}$. If $x_0=0.9$ and $D=1$, then $v_c^*\approx 2.22$. Again we find similar behavior for $z>2$ and all the graphs asymptotically approach each other as $v\rightarrow \infty$. In addition, we have the ordering
 \begin{equation}
 \label{bound}
 v_c^{*}(z)<v_c^{**}(z) <v_c(z)
 \end{equation}
 for all coordination numbers $z$. One subtle point concerns how the LD transition manifests itself in the presence of resetting, given that $\pi_r(x_0)=1$ and $T_r(x_0)<\infty$ for all $r>0$ and $v$. In particular,
\begin{equation}
\lim_{r\rightarrow 0^+}\pi_r(x_0)=1 \mbox{ for all } v.
\end{equation}
On the other hand, $\pi_0(x_0) <1 $ and thus $T_0(x_0)=\infty$ for $v>v_c$. In Fig. \ref{fig9} we plot the MFPT $T_r(x_0)$ as a function of $v$ for a decreasing set of resetting rates $r$. As $r\rightarrow 0^+$, one finds that there is a rapid increase in $T_r$ as $v$ crosses the critical velocity $v_c(z)$ from below, beyond which $T_r$ flattens out. This rapid increase is the remnant or ghost of the LD transition without resetting.

\section{Quenched disorder} 

So far we have focused on homogeneous Cayley trees. However, in a number of applications one finds that the transport properties are randomly distributed across the network, and the resulting quenched disorder can lead to anomalous behavior, fractal scaling and
critical phenomena \cite{Havlin,Georges}. The effects of quenched disorder on the steady-state solution at a closed terminal node (without resetting) and the LD transition were considered in \cite{Bressloff96,Bressloff97}. In order to combine the effects of quenched disorder and stochastic resetting, it is necessary to consider the full solution in Laplace space. In this section we first summarize some of the results of our previous work, and then show how the inclusion of stochastic resetting naturally leads to the theory of random matrix products. This is a topic of considerable interest within probability theory and statistical mechanics \cite{Wishart28,Kesten60,Kesten73,Derrida83,Crisanti93,Burda12,Forrester14}.

\subsection{Steady-state solution}
Following Ref. \cite{Bressloff96}, suppose that for each generation
$n$, all branches $i\in \Gamma$ such that $\alpha(i)=n$ have the same drift velocity
$ v_n$ but the sequence $\{v_n, n \geq 0\}$ is independently and identically distributed (inter-generational quenched disorder). 
Let $\rho(v)$ be the
associated probability density. The density evolves according
to the inhomogeneous FP equation
\begin{eqnarray}
\label{FPq}
  \frac{\partial p_i}{\partial t} = D\frac{\partial^2 p_i}{\partial
    x^2}-v_n\frac{\partial p_i}{\partial x} ,
  \quad 0 < x < L
\end{eqnarray} 
for all $i$ such that $\alpha(i)\in \Sigma_n$.
The corresponding probability current or flux on the branch is now
\begin{equation}
  \flx_n[p]\equiv  - D\frac{\partial p}
        {\partial x} +v_n p.
\end{equation}
In the presence of inter-generational quenched disorder, equation (\ref{A}) becomes
\begin{eqnarray}
A_0^{-1}\equiv Y = f(v_0)+\sum_{n=1}^{\infty}f(v_n)(z-1)^{n}\prod_{m=0}^{n-1}\bar{f}(v_m),
\label{AA}
\end{eqnarray}
where 
\begin{equation}
f(v)=\frac{D}{v}(\mbox{e}^{vL/D}-1),\quad \bar{f}(v)=\mbox{e}^{vL/D}.
\end{equation} 
We assume that $v$ is finite so $f(v),\bar{f}(v)$ are bounded, positive functions. 
The steady-state density is thus expressed in terms of a random geometric series $Y$. Similar series have arisen in a variety of
studies of one-dimensional problems \cite{Georges,Kesten75,Solomon75,Calan85}.
It can be proven that if
\begin{equation}
\langle \ln[(z-1)\bar{f}(v)] \rangle < 0,
\end{equation}
 then $Y$ converges with probability one \cite{Kesten73}. Hence, the steady-state is localized
provided that
\begin{equation}
\langle v \rangle  < v_c(z) =-\frac{D}{L}\ln (z-1), 
\end{equation}
that is, the average drift velocity is less than the critical velocity for localization on a homogeneous
tree. On the other hand, if $\langle v \rangle > v_c(z)$ then $Y$ is infinite and the steady-state is delocalized. The critical point determines a phase boundary in the infinite-dimensional space of probability densities $\rho(v)$ that
separates the localized and delocalized phases.  A characteristic feature of the phase transition is that as
$\langle v \rangle 
\rightarrow v_c$ in some prescribed fashion, the probability distribution $\F$ of $Y$ in the localized phase develops a long-tail for which all
moments are infinite, even though the system is still localized. This is a consequence of Jensen's inequality 
\begin{equation}
\langle \mbox{e}^{vL/D}\rangle \geq \mbox{e}^{\langle v\rangle L/D},
\end{equation}
 which means that one can have $\langle v \rangle < v_c$ even though $\langle \mbox{e}^{vL/D}\rangle=\e^{Lv_c/D}$. 

The geometric series (\ref{AA}) can be generated iteratively using the difference equation
\begin{equation}
\label{YN}
Y^{(N)}_{n}=(z-1)\bar{f}(v_n)Y^{(N)}_{n+1}+f(v_n),\quad 0\leq n \leq N -1
\end{equation}
with $Y^{(N)}_N$ fixed and each pair $(f(v_n),\bar{f}(v_n))$ generated independently from $\rho(v)$. Introducing the probability density
\begin{equation}
P^{(N)}_n(y)dy=\P[y<Y_n^{(N)}<y+dy],
\end{equation}
we have the integral equation
\begin{eqnarray}
\label{Pin}
P_n^{(N)}(y)&=\int_{-\infty}^{\infty}dv \rho(v)\, \int_{-\infty}^{\infty}dw \, \delta(y-(z-1)\bar{f}(v)w-f(v))P_{n+1}^{(N)}(w)\nonumber \\
&=\int_{-\infty}^{\infty}  \rho(v)\,\frac{\rho(v)}{(z-1)\bar{f}(v)}P_{n+1}^{(N)}\left (\frac{y-f(v)}{(z-1)\bar{f}(v)}\right )dv. 
\end{eqnarray}
Taking the limit $N\rightarrow \infty$ on both sides for fixed $n$ with $P_n^{(N)}(y)\rightarrow \Psi(y)={\mathcal F}'(y)$ yields the following integral equation for
$\Psi(y)$:
\begin{eqnarray}
\label{int1}
\Psi(y)
&=\int_{-\infty}^{\infty} \rho(v)\,\frac{\rho(v)}{(z-1)\bar{f}(v)}\Psi\left (\frac{y-f(v)}{(z-1)\bar{f}(v)}\right )dv.
\end{eqnarray}
An alternative form of the integral equation is obtained by taking Laplace
transforms of the first line in equation (\ref{Pin}),
\begin{eqnarray}
\label{int2}
M(s)=\int_{-\infty}^{\infty}\rho(v) M((z-1)\bar{f}(v)s)\mbox{e}^{-sf(v)}dv
\end{eqnarray}
with $M(s) =\int_0^{\infty}\mbox{e}^{-sy}\Psi(y)dy$.

It is not generally possible to solve these integral equations analytically. However, one can determine the
asymptotic behavior of
$\Psi$ when $y$ is large. We illustrate this for the Gaussian distribution
\begin{eqnarray}
\label{gauss}
\rho(v)=\frac{1}{\sqrt{2\pi \Delta^2}}\exp \left (-\frac{(v-\mu)^2}{2\Delta^2} \right ).
\end{eqnarray}
(The theory for more general distributions is considered in \cite{Bressloff97}.)
 Suppose that $\langle v \rangle  < v_c(z)$ but the first moment of $\Psi$ is
infinite so that
$\langle \bar{f}(v)\rangle >( z-1)^{-1}$. It can then be proven \cite{Kesten75} that there exist positive
constants $a,b,\sigma$ with $0 < \sigma < 1$ such that 
$\Psi(y)
\sim ay^{-\sigma -1}$
for large $y$, and hence
$M(s) \sim 1 + bs^{\sigma}$
for small $s$. The large--$y$ behavior of $\Psi$ ensures that if $\sigma > 0$ then
\begin{eqnarray}
{\mathcal F}^*\equiv \lim_{x
\rightarrow \infty}\int_x^{\infty}\Psi(y)dy=0.
\end{eqnarray}
That is, the series $Y$ of equation (\ref{AA}) is
convergent with probability one. 
Substitution of the
asymptotic form for
$\Psi$ (or $M$) into equation (\ref{int1}) (or (\ref{int2})) leads to the equation
\begin{eqnarray}
\fl 1&=\beta(\sigma)\equiv (z-1)^{\sigma}\langle
 \bar{f}(v)^{\sigma}
\rangle =(z-1)^{\sigma}\exp \left (\mu \sigma L/D+\sigma^2 \Delta^2 L^2/2D^2 \right ).
\end{eqnarray}
where the last line follows from plugging in the Gaussian distribution. Note that $\beta(0) = 1$ and $\beta(\sigma)$ is a convex function for real
$\sigma$. For $\mu < 0$, there exists a second solution 
$\beta(\sigma^*)=1$ with
\begin{equation}
 \sigma^* = (v_c(z)-\mu)\frac{2D}{L\Delta^2} .
 \end{equation}
 As $\mu \equiv \langle v\rangle \rightarrow v_c(z)$ from below we see that $ \sigma^*(\mu)\rightarrow 0$ and $\Psi(y)$ is no longer normalizable, signaling  an LD transition.
 
 \subsection{Solution in Laplace space}

In order to combine the effects of quenched disorder and stochastic resetting, it is necessary to consider the full solution in Laplace space. Equation (\ref{AV}) becomes
\begin{equation}
\label{qAV}
\left [D\frac{\partial^{2} }{\partial x^{2}} 
     -v_n\frac{\partial}{\partial x} -s \right ]\tu_{i}(x,s) = - \delta_{i,0}\delta(x-x_0),\quad  \alpha(i)\in \Sigma_n.
\end{equation}
The corresponding solution (\ref{GB}) generalizes as
  \begin{eqnarray}
  \label{qGB}
\fl   \tu_i(x,s) &= \delta_{i,0}{\mathcal G}_0(x,x_0;s) +
    \widetilde{\Phi}_{n-1}(s)\frac{\psi_n(x-L,s)}{\psi_n(-L,s)} + \widetilde{\Phi}_{n}(s)\frac{\psi_n(x,s)}{\psi_n(L,s)},\ \alpha(i)\in \Sigma_n,
  \end{eqnarray}
 with
\begin{eqnarray}
\label{qbag}
 \fl \uoo_n(x,s) &= \e^{v_nx/2D} \left [\e^{\eta_n(s)x} - \e^{-\eta_n(s)x}\right ],\quad \eta_n(s)=\frac{\sqrt{v_n^2+4Ds}}{2D}.
  \end{eqnarray}
  The Green's function ${\mathcal G}_0$ is given by equation (\ref{G}) with $v=v_0$, where $v_0$ is the velocity on the primary branch. Proceeding along similar lines to the homogeneous case, we impose current conservation at each branch node. We thus obtain modified versions of equations (\ref{pip}) and (\ref{itp}) given by
\begin{equation}
\label{qpip}
  g_n\widetilde{\Phi}_{n-1}+h_n\widetilde{\Phi}_{n}
  -(z-1)[ \bar{h}_{n+1} \widetilde{\Phi}_{n}
 +\bar{g}_{n+1}\widetilde{\Phi}_{n+1} ]=   \chi_0\delta_{n,0}
\end{equation}
for $n\geq0$ and $\targT_0\equiv - \flx[{\mathcal G}_0](L)$.
The functions $g_n,h_n,\bar{g}_n,\bar{h}_n$ are given by equations (\ref{gs})--(\ref{hs}) except that $v\rightarrow v_n$. For example, making the velocity dependence of $g$ explicit, we have $g_n(s)=g(s,v_n)$, and similarly for the other variables.

Equation (\ref{qpip}) is a second-order analog of the random difference equation (\ref{YN}). It is convenient to rewrite it as a first-order matrix equation by defining a second variable $\widetilde{\Psi}_{n+1}=\widetilde{\Phi}_{n+1}-\widetilde{\Phi}_n$ such that
\begin{equation}
\label{matrix}
(z-1)\overline{\Lambda}_{n+1}(s)\left (\begin{array}{c}\widetilde{\Phi}_{n+1}\\ \widetilde{\Psi}_{n+1}\end{array}\right ) =\Lambda_n(s) \left (\begin{array}{c}\widetilde{\Phi}_{n}\\ \widetilde{\Psi}_{n}\end{array}\right )-\delta_{n,0}\left (\begin{array}{c}\chi_0(s)\\ 0\end{array}\right )
\end{equation}
with
\begin{equation}
\fl \Lambda_n(s)=\left(\begin{array}{cc}  {h}_{n}(s)+  g_{n}(s) & - g_{n}(s)\\
z-1 &0 \end{array}\right ),\overline{\Lambda}_n(s)=\left (\begin{array}{cc} \bar{h}_{n}(s)+ \bar{g}_{n}(s) & - \bar{h}_{n}(s)\\
1 &-1 \end{array}\right ).
 \end{equation} 
 Iterating equation (\ref{matrix}) gives
 \begin{eqnarray}
 \label{prod}
\fl  (z-1) \overline{\Lambda}_N (s)\left (\begin{array}{c}\widetilde{\Phi}_{N}\\ \widetilde{\Psi}_{N}\end{array}\right )= \left [\prod_{m=1}^{N-1} \Gamma_m(s)\right ]\left [\Lambda_0(s) \left (\begin{array}{c}\widetilde{\Phi}_{0}\\ \widetilde{\Psi}_{0}\end{array}\right )-\left (\begin{array}{c}\chi_0(s)\\ 0\end{array}\right )\right ],
 \end{eqnarray}
 where
 \begin{eqnarray}
  \Gamma_n(s)&=\frac{1}{z-1}\Lambda_n(s)\overline{\Lambda}_n(s)^{-1}\nonumber \\
  &=\frac{1}{(z-1)\overline{g}_n(s)} \left (\begin{array}{cc} h_n(s) & g_n(s)\bar{g}_{n}(s)- h_n(r)\bar{h}_{n}(s)\\
z-1 &-(z-1)\bar{h}_n(s) \end{array}\right ).
\label{Gam}
\end{eqnarray}
Before proceeding further, it is useful to check that the solution for a homogeneous Cayley tree is recovered when $v_n=v$ for all $n$. Dropping the index $n$ in equation (\ref{Gam}), we find that the eigenvalues of the matrix $\Gamma(s)$ are $\lambda_{\pm}(s)$ with $\lambda_{\pm}(s)$ defined in equation (\ref{lam0}). The corresponding eigenvectors (up to scalar multiplication) are 
\begin{equation}
{\bf c}_{\pm}(s)=\left (\begin{array}{c} (z-1)\bar{h}(s)+\lambda_{\pm}(s) \\ z-1  \end{array} \right ).
\end{equation}
Recall that normalizability of the solution requires that $\widetilde{\Phi}_N$ decays faster than $(z-1)^{-N}$, which means that the eigenvector on the right-hand side of equation (\ref{prod}) should project onto the eigenvector of the smaller eigenvalue $\lambda_-(s)$. In other words,
\begin{equation}
\Lambda_0(s) \left (\begin{array}{c}\widetilde{\Phi}_{0}(s)\\ \widetilde{\Psi}_{0}(s)\end{array}\right )-\left (\begin{array}{c}\chi_0(s)\\ 0\end{array}\right )\propto {\bf c}_-(s).
\end{equation}
Taking ratios of the components on both sides and rearranging leads to equation (\ref{phi0}), and the analytical results regarding phase transitions in a homogeneous Cayley tree are recovered.

\begin{figure}[b!]
\raggedleft
\includegraphics[width=13cm]{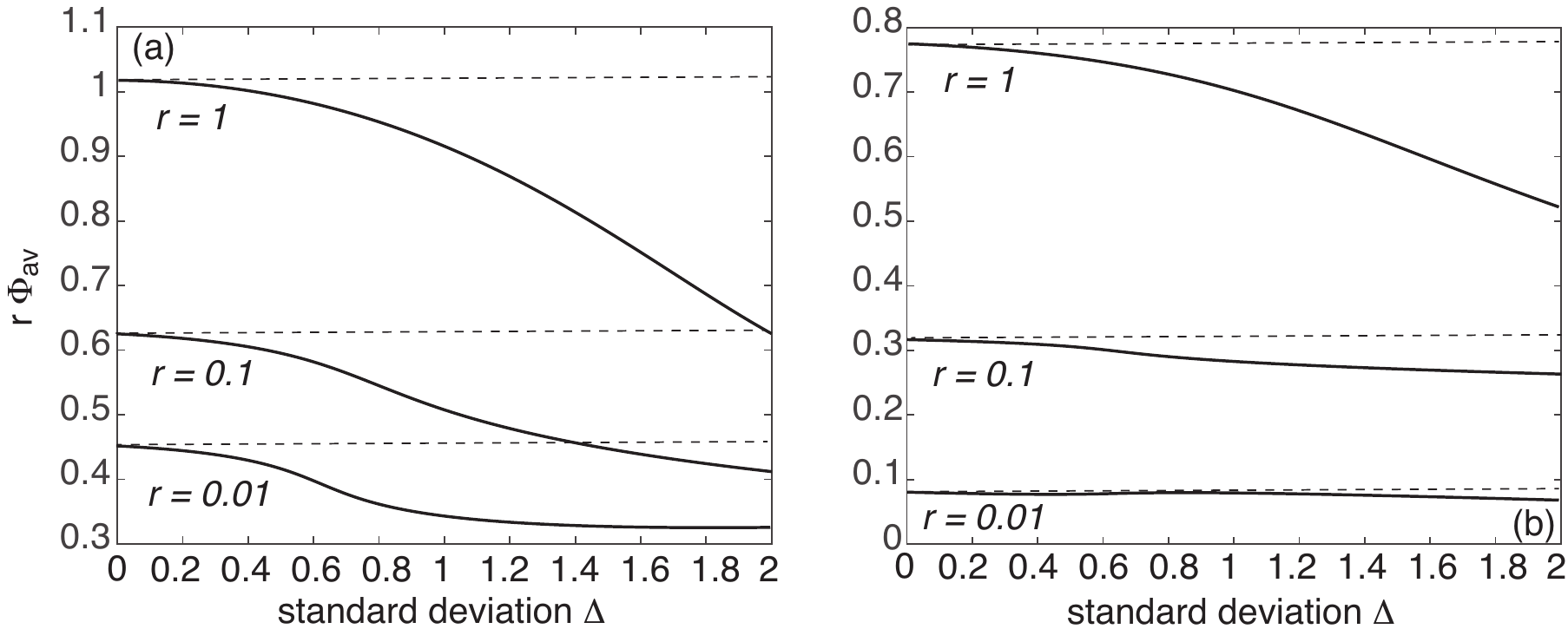} 
\caption{Effective steady state $r\Phi_{\rm av}(x_0,r)$ at the terminal node for the averaged system with stochastic resetting at a rate $r$. The branch velocities are distributed according to the Gaussian (\ref{gauss}) with standard deviation $\Delta$ and mean $\mu$. (a) Plot of the steady state as a function of $\Delta$ and various resetting rates for $\mu=-1$. (b) Corresponding plots for $\mu=-0.5$. Horizontal dashed lines indicate the steady-state solutions for a homogeneous tree with branch velocity $v=\mu$. Other parameters are $z=3$, $x_0=0.5$, $D=1$ and $L=1$.}
\label{fig11}
\end{figure}

Incorporating the effects of quenched disorder requires determining the statistical properties of the random matrix product 
\begin{equation}
\label{Thet}
\Theta_{N}(s)=\prod_{m=1}^{N-1} \Gamma_m(s).
\end{equation}
 The analysis is complicated by the fact that the entries of each independent matrix $\Gamma_n$ are statistically correlated through their dependence on $v_n$, rather than being independent identically distributed random variables. Moreover, $\Gamma_n$ is non-symmetric and not all components are positive. Hence, classical results concerning random matrix products do not easily apply \cite{Wishart28,Kesten60,Kesten73,Derrida83,Crisanti93,Burda12,Forrester14}.
Motivated by the analysis of the homogeneous case, we restrict ourselves to solving a simpler problem in which $\Theta_{N}(s)$ is replaced by the matrix product obtained by averaging with respect to the independent random velocities:
\begin{equation}
\label{Thetav}
\langle \Theta_{N}(s)\rangle =\prod_{m=1}^{N-1}\langle \Gamma_m(s)\rangle =\Gamma(s)^{N-1},
\end{equation}
where
\begin{eqnarray}
 \fl  \Gamma(s) =\frac{1}{(z-1)} \left (\begin{array}{cc} \langle h(s)/\overline{g}(s)\rangle & \langle g(s)\rangle- \langle h (s)\bar{h}(s)/\overline{g}(s)\rangle\\
(z-1)\langle 1/\bar{g}(s)\rangle  &-(z-1)\langle \bar{h}(s)/\bar{g}(s)\rangle \end{array}\right ).
\end{eqnarray} 
Here
\[\langle g(s)\rangle =\int_{-\infty}^{\infty}\rho(v_n)g(s,v_n)dv_n\]
etc. As a further simplification, suppose that $v_0=\langle v\rangle$. Let $\Phi_{\rm av}(x_0,s)$ denote the steady-state solution at the terminal node. Following the arguments of the homogeneous case, $\Phi_{\rm av}(x_0,s)$ takes the form (\ref{Phi1}) with $\lambda$ now the smaller eigenvalue of the averaged matrix $\Gamma(s)$:
\begin{eqnarray}
\lambda=\frac{w(s)- \sqrt{w(s)^2+4\hat{w}(s)}}{2(z-1)} 
\end{eqnarray}
with
\begin{equation}
w(s)=\left \langle \frac{{h}(s)}{\bar{g}(s)}\right \rangle-(z-1)\left \langle \frac{\bar{h}(s)}{\bar{g}(s)}\right \rangle>0
\end{equation}
and
\begin{equation}
\fl \hat{w}(s)= (z-1)\left [\left \{\langle g(s)\rangle- \left \langle \frac{ h (s)\bar{h}(s)}{\overline{g}(s)}\right \rangle\right \}\left \langle \frac{ 1}{\overline{g}(s)}\right \rangle+\left \langle \frac{ h (s)}{\overline{g}(s)}\right \rangle\left \langle \frac{ \bar{h}(s)}{\overline{g}(s)}\right \rangle\right ].
\end{equation}
Although one cannot simply identify $\Phi_{\rm av}(x_0,s)$ with the first moment $\langle \widetilde{\Phi}_{0}(s)\rangle$, since the vector on the right-hand side of equation (\ref{prod}) is not statistically independent of the matrix $\Theta_N$, it can be checked that the first moment localization condition $\langle \mbox{e}^{vL/D}\rangle=\e^{Lv_c/D}$ is recovered in the limit $s\rightarrow 0$. This follows from the identities (\ref{ids2}), which imply
\begin{equation}
 w(0)= (z-1)\langle \e^{vL/D}\rangle+1,\quad \hat{w}(0)=-(z-1)\langle \e^{vL/D}\rangle 
\end{equation}
and
\begin{equation}
\lambda(0)=\frac{(z-1)\langle \e^{vL/D}\rangle+1-| (z-1)\langle \e^{vL/D}\rangle-1|}{2(z-1)}.
\end{equation}
Hence, if $(z-1)\langle \e^{vL/D}\rangle <0$ then $\lambda_-(0)=\langle \e^{vL/D}\rangle$, whereas $\lambda_-(0) =1/(z-1)$ when $(z-1)\langle \e^{vL/D}\rangle >0$. 

The quantity $r\Phi_{\rm av}(x_0,r)$ determines the steady-state solution at the closed terminal node for the averaged system. For the sake of illustration, suppose that $\rho(v)$ is given by the Gaussian (\ref{gauss}). In Fig. \ref{fig11} we plot $r\Phi_{\rm av}(x_0,r)$  as a function of the standard deviation $\Delta$ for fixed means $\mu$ and various resetting rates. It can be seen that increasing the variance of the quenched disorder reduces the steady state.

\setcounter{equation}{0}
 \section{Discussion}

In summary, we identified two distinct examples of phase transitions in the optimal resetting rate for drift-diffusion on a semi-infinite Cayley tree, one associated with maxima of the steady-state at a closed terminal node and the other other associated with minima of the MFPT to reach an open terminal node. In both cases, the critical velocities are bounded from above by the critical velocity of the LD transition without resetting, see equation (\ref{bound}). The latter involves a phase transition between zero and nonzero values of the steady-state at a closed terminal node or, equivalently, the transition between being absorbed by an open terminal node with probability one or with probability less than one. Only the critical velocity $v_c(z)$ for the LD transition appears to have a simple universal dependence on $z$, namely, $v_c(z)=-D\ln(z-1)/L$ for $z>2$ and $v_c(2)=0$. All three phase transitions arise from the same basic principle, namely, that there is a crossover between a diffusion-dominated and a drift-dominated regime. In the presence of stochastic resetting, this crossover point is shifted towards more negative drift velocities since the reset point is away from the terminal node and thus resetting contributes an effective positive velocity component to the transport process. Finally note that all three phase transitions are second-order, since there is a continuous change in the relevant ``order parameter.'' (Examples of first-order phase transitions in optimal resetting rates are considered in \cite{Pal19}.)

For the sake of analytical tractability, we assumed that the reset point of the particle was located on the primary branch so that all branches of a given generation could be treated the same. Such a simplification would also hold if the particle randomly reset to one of the nodes of the $n$-th generation with probability $2^{-n}$. Although the analysis would be less straightforward under more general choices of resetting protocol, the basic conclusions of the paper would not be altered. In particular, the LD transition is a property of the underlying geometry as expressed by the coordination number $z$.
In the final part of the paper we considered the case of quenched disorder in the branch velocities, and showed how the inclusion of stochastic resetting naturally leads to a problem in the theory of random matrix products. In future work it would be interesting to use this theory to go beyond the naive calculation carried out for the averaged system.

\end{document}